\documentclass[12pt,draftclsnofoot, ,onecolumn]{IEEEtran}
\usepackage{algorithm,algorithmic}
\usepackage[subtle]{savetrees}
\usepackage{verbatim}
\usepackage{graphicx}
\usepackage{subcaption}
\usepackage{amsmath,amssymb,amsfonts,bm}
\usepackage{epsf}
\usepackage{tikz}
\usepackage{psfrag}
\usepackage{pstool}
\usepackage{epstopdf}
\usepackage{booktabs}
\usepackage{stfloats}
\usepackage{url}
\usepackage[dvips]{epsfig}
\usepackage{amsthm}
\usepackage{color}
\usepackage[usenames,dvipsnames]{pstricks}
\usepackage{lettrine}
\usepackage[dvips]{epsfig}
\usepackage{pst-grad} % For gradients
\usepackage{pst-plot} % For axes
\usepackage{epsfig}
\usepackage{enumerate}
\usepackage{amsbsy}
\usepackage{amssymb}
\usepackage{amsthm}
\usepackage{amscd}
\usepackage{float}
\usepackage{color}
\usepackage[numbers,sort&compress,square]{natbib}

\usepackage{listings}
\usepackage{stackrel}
\usepackage{authblk}
\usepackage{dsfont}
\usepackage{ulem}
\makeatletter
\newcommand{\removelatexerror}{\let\@latex@error\@gobble}
\makeatother
\newtheorem{thm}{Theorem}

\newtheorem{assmp}{Assumption}

\usepackage{stackengine}
\def\delequal{\mathrel{\ensurestackMath{\stackon[1pt]{=}{\scriptstyle\Delta}}}}
\definecolor{mintbg}{rgb}{.63,.79,.95}
\usepackage{enumitem}
\begin{document}
%\title{Federated Learning Over HAPS Networks: A Delay-Accuracy Trade-Off Problem}
\title{FLSTRA: Federated Learning in Stratosphere}
\author{
Amin Farajzadeh, Animesh Yadav, Omid Abbasi, Wael Jaafar, and Halim Yanikomeroglu
\thanks{A. Farajzadeh, A. Yadav, O. Abbasi, and H. Yanikomeroglu are with the Non-Terrestrial Networks (NTN) Lab, Department of Systems and Computer Engineering, Carleton University, Ottawa, ON K1S 5B6, Canada (e-mails: \{aminfarajzadeh, animeshyadav, omidabbasi, halim\}@sce.carleton.ca).\\
\indent Wael Jaafar is with the department of Software and IT Engineering, École de technologie supérieure (ÉTS), Montreal, QC H3C 1K3, Canada (e-mail: wael.jaafar@etsmtl.ca).}}

\makeatletter
\patchcmd{\@maketitle}
  {\addvspace{0.5\baselineskip}\egroup}
  {\addvspace{-1\baselineskip}\egroup}
  {}
  {}
\makeatother
\maketitle
\begin{abstract}
% Start directly with FL challange in terrestrial ....
% then we proposing HAPS...
% trade-off
%FL over wireless (terr)

We propose a federated learning (FL) in stratosphere (FLSTRA) system, where a high altitude platform station (HAPS) facilitates a large number of terrestrial clients to collaboratively learn a global model without sharing the training data. FLSTRA overcomes the challenges faced by FL in terrestrial networks, such as slow convergence and high communication delay due to limited client participation and multi-hop communications. HAPS leverages its altitude and size to allow the participation of more clients with line-of-sight (LOS) links and the placement of a powerful server. However, handling many clients at once introduces computing and transmission delays. Thus, we aim to obtain a delay-accuracy trade-off for FLSTRA. Specifically, we first develop a joint client selection and resource allocation algorithm for uplink and downlink to minimize the FL delay subject to the energy and quality-of-service (QoS) constraints. Second, we propose a communication and computation resource-aware (CCRA-FL) algorithm to achieve the target FL accuracy while deriving an upper bound for its convergence rate. The formulated problem is non-convex; thus, we propose an iterative algorithm to solve it. Simulation results demonstrate the effectiveness of the proposed FLSTRA system, compared to terrestrial benchmarks, in terms of FL delay and accuracy.
%We propose a federated learning (FL) in stratosphere (FLSTRA) system, where a high altitude platform station (HAPS) felicitates a large number of terrestrial clients to learn a global model by sharing their local updates while preserving local data privacy. FLSTRA overcomes the challenges faced by FL in terrestrial networks, such as slow convergence and high communication delay due to limited user participation and multi-hop communications. HAPS leverages its altitude and size to allow the participation of a larger number of clients, each with line-of-sight (LoS) links and the placement of a powerful server. However, handling many clients at once introduces computing and transmission delays. Thus, we aim to obtain a delay-accuracy trade-off for FLSTRA. Specifically, we first develop a joint client selection and resource allocation algorithm for uplink and downlink to minimize the FL delay subject to the energy and quality-of-service (QoS) constraints. Second, we propose a communication and computation resource-aware (CCRA-FL) algorithm to achieve the target FL accuracy while developing an upper bound for its convergence rate. The formulated problem is non-convex; thus, we propose an iterative algorithm to solve it. Obtained simulation results demonstrate the effectiveness of the proposed idea of utilizing a HAPS server for FL systems, compared to terrestrial benchmarks, in terms of FL delay and accuracy.
%The convergence behaviour of FL with HAPS is also investigated and simulation results reveal better performance compared to that of terrestrial FL systems. 
\end{abstract}
%%%%%%%%%%%%%%%%%%%%
\section{Introduction}
\lettrine[]{R}{apid} social changes are reshaping individuals' and communities' requirements and expectations regarding network services, use-cases, and key performance indicators (KPIs)~\cite{intro1}. A highly adaptive, real-time, intelligent, privacy-preserving, and self-evolving mechanism seems necessary for the next generation of wireless networks to respond to these varying demands and requirements~\cite{intro2}.
%This requires beyond 5G networks to be more of distributed, decentralized, and intelligent innovative network. To this direction, the main pillar is the twofold use of artificial intelligence (AI)/machine learning (ML), both as a means to efficiently orchestrate the wireless networks and as the core of decentralizing the network service architecture.
The decentralized machine learning (ML) framework has recently gained attention as an efficient solution to enable end-to-end intelligent and real-time decision-making in wireless networks~\cite{intro3}. In particular, the federated learning (FL) technique, developed by Google in 2016~\cite{intro7}, provides distributed intelligence for decentralized model training. The basic idea of FL is to invite the network's active users with sufficient computation capability, i.e., equipped with computing modules such as CPU, GPU, and RAM, and data to participate in predicting or estimating network parameters, i.e., model training~\cite{asyn1}. FL has an inherent privacy-preserving mechanism that guarantees the data privacy of network users~\cite{intro4}. Indeed, it allows collaborative model training without explicitly sharing local data. In FL, each user shares its model training parameters with a centralized server, which aggregates model training parameters after receiving them from all the users and broadcast the aggregated parameter back to the users. This cycle continues for several rounds (also known as the communication round) until convergence is achieved~\cite{surv3}.
%However, there exist some challenges that degrade its performance from delay and accuracy perspectives. 
\vspace{-4mm}
\subsection{Related Works}

FL over terrestrial wireless networks has gained considerable attention in recent years. Researchers have been exploring various approaches to improve the performance of FL over wireless networks. Several research studies have been conducted to improve the performance of FL over wireless networks. For instance, in~\cite{RW1}, the authors proposed an asynchronous FL algorithm that selects a subset of clients based on their arrival order to be aggregated at the FL server, thereby improving communication efficiency. In~\cite{fl1}, the authors investigate energy-efficient transmission and computation resource allocation for FL over wireless communication networks. They formulate the joint learning and communication problem as an optimization problem to minimize energy consumption under a latency constraint and propose an iterative algorithm to solve it.

Moreover, the authors in~\cite{swipt} proposed a simultaneous wireless information and power transfer (SWIPT) aided FL, in which an FL server simultaneously broadcasts the global model and provides wireless power transfer to wireless devices, thereby improving the energy efficiency of the system. In~\cite{assumps}, the authors considered FL with multiple edge-servers and accelerated the training by utilizing the clients located in the overlapping areas among different edge-servers. The authors in~\cite{HFL} proposed a novel hierarchical FL framework where model aggregation can be partially migrated from the cloud to edge-servers.
\vspace{-4mm}
 % add FL over wireless terr for smooth transition to next phara -- refereneces
\subsection{Motivations and Contributions}
The existing methods discussed have demonstrated significant improvements in the performance of FL over terrestrial wireless networks. Nevertheless, certain challenges still persist, including limitations on the number of participants, heterogeneous wireless environments, non-line-of-sight (NLOS) channel failures, and heterogeneous client computing resources. Research continues to address these challenges. In terrestrial networks, due to limited coverage and network resources such as bandwidth, client availability instability, and channel failures, only a portion of clients can participate in each FL round reliably~\cite{intro4B}. Considering the iterative nature of FL, limiting the number of participants often results in a high computation delay and low training accuracy~\cite{intro5}. Moreover, for an FL system in terrestrial settings usually, multiple base stations (BSs) are employed as a relay to communicate with the server in the cloud server, and since the raw data is distributed across the edge devices in the access network, there are multi-hop wireless links thereby FL suffers from inevitable high communication delay. FL delay issues are studied in current literature~\cite{intro6}, and the efficiency of existing approaches degrades with the communication delay. Hence, stringent delay requirements render the centralized configurations of FL in terrestrial networks impractical for forthcoming applications, such as smart grids, autonomous vehicles, e-health, and augmented reality~\cite{intro8}. 

In order to address the challenges associated with traditional network architectures, innovative designs that can provide improved coverage and more reliable communication channels are required. One potential solution is the use of non-terrestrial platforms for FL, which can offer a more efficient and dependable means of communication in situations where traditional terrestrial base stations are limited in their coverage or communication reliability. However, it is important to note that the selection of an appropriate platform will depend on a variety of factors, and non-terrestrial platforms may not be suitable or feasible for all scenarios. With the limited coverage and environment-dependent communication reliability offered by terrestrial base stations, the use of a non-terrestrial platform like the  high altitude platform station (HAPS) seems promising~\cite{haps1}. In fact, HAPS has recently been proposed as a potential candidate for 6G networks, which aim to provide ubiquitous communication~\cite{HAPS1}. HAPS can operate at the stratospheric altitude of around $20$ km, providing strong line-of-sight (LOS) communications and wide coverage, with a radius ranging between $50$ and $500$ km~\cite{intro-haps2}. HAPS has a large payload, which allows for hosting robust communication, computing and storage equipment, and long-lasting batteries~\cite{HAPS2}. Given these advantages, HAPS can reliably communicate with a massive number of users and involve them as participants in the FL process. 
%while minimizing the channel failures. 
In addition, it is expected that HAPS-enabled FL can achieve faster convergence and better model accuracy than its terrestrial counterparts due to the improvement in wireless channel quality and the participation of a large number of users in the FL training process. Recently, in~\cite{FedHAP}, the strategic position of HAPS is leveraged for fast and efficient FL for low earth orbit (LEO) satellite constellations. Essentially, HAPS is introduced to facilitate the global aggregation of local models received by high-speed orbiting satellite clients. In contrast, our work focuses on an FL system that uses a single HAPS as a server to facilitate collaborative learning among a large number of terrestrial clients. As a result, our approach is designed to cater to a larger number of terrestrial clients, while FedHAP is designed to cater to satellites that are moving. This fundamental difference in the network architecture leads to distinct design and optimization challenges, which we address in our work.

In fact, the integration of FL into HAPS networks can offer potential use cases. For example, in the smart city context, HAPS can be utilized for traffic management and surveillance, where FL models can be trained on data collected from sensors and cameras across the city. Autonomous vehicles and drones can also benefit from FL over HAPS networks for real-time decision-making based on local data. In the healthcare sector, FL over HAPS networks can be utilized for remote patient monitoring and diagnosis, where privacy-preserving FL models can be trained on patient data collected from wearable devices distributed in wide geographical regions and sent to HAPS for aggregation. In augmented reality and virtual reality applications, FL over HAPS networks can enable real-time processing of large amounts of data collected from sensors and cameras to create large-scale immersive experiences. Moreover, FL over HAPS networks has the potential to provide more efficient and reliable solutions for various use cases. In the case of autonomous vehicles, the use of HAPS networks can facilitate better hand-off between base stations, enabling more efficient and reliable communication during the hand-off process. Furthermore, HAPS networks can provide a platform for large-scale FL systems over wide geographical scales, which is especially relevant for use cases such as disaster response, where a large number of participants may be spread over a wide area.

On the other hand, non-independent and identically distributed (non-IID) data distributions can hinder convergence speed in FL and negatively impact accuracy even with a large number of participants~\cite{intro-fl}. To address this issue, a novel mechanism is required to control the number of selected clients based on their data distributions. For instance, in healthcare applications, different hospitals may have vastly different patient demographics and medical conditions, leading to non-IID data distributions~\cite{medical}. Our proposed mechanism optimizes the selection of clients to achieve the best possible balance between FL delay/convergence speed and accuracy, even in highly non-IID scenarios. By carefully controlling the number of selected clients, the accuracy of the FL model can be significantly improved while minimizing the end-to-end FL delay. Our study is the first to propose a joint participant selection and resource allocation design strategy in the context of a HAPS-enabled FL system, considering both delay and accuracy perspectives.
The key contributions of this paper can be summarized as follows:
\begin{itemize}
\item We discuss the current challenges of FL in terrestrial settings, such as limited coverage, heterogeneous wireless environments, and unstable client availability, and highlight the unique benefits that HAPS can provide for FL to solve them. HAPS, with its large coverage area and direct line-of-sight links, can mitigate these challenges and enable better FL performance.
\item We propose a novel resource-aware FL algorithm that achieves the desired FL accuracy while minimizing communication and computation costs. We also derive an upper bound for the convergence rate of the proposed FL algorithm as a function of the number of selected clients, communication rounds, and location iterations. This algorithm can help to overcome the limited computing resources of wireless clients and reduce communication costs.
\item We formulate a joint communication and computation optimization problem to minimize the end-to-end FL delay for both uplink and downlink. To solve this problem, we propose an iterative algorithm and obtain closed-form solutions. This optimization problem considers the resources of both HAPS and clients and aims to achieve a balance between FL accuracy and delay.
\item Our simulation results demonstrate that HAPS can significantly improve FL performance compared to terrestrial networks. The large footprint and line-of-sight links of HAPS are advantageous for FL systems, and our proposed scheme that jointly optimizes HAPS and clients' resources can reduce the FL delay while providing a trade-off between FL accuracy and delay. Additionally, we show that the undesirable displacement of HAPS has a negligible effect on the FL delay performance.
\end{itemize}
\vspace{-6mm}
\subsection{Organization}
\indent The rest of this paper is organized as follows. In Section~\ref{sys}, the system model is described and a communication and computation resource-aware FL algorithm is developed. Section~\ref{prob} formulates the related optimization problem and exposes the proposed iterative algorithm for end-to-end FL delay minimization. Then, simulation results are presented and discussed in Section~\ref{sim}. Finally, the conclusion is drawn in Section~\ref{concl}.

 %%%%%%%%
 \begin{table*}[!t] \label{notation}
 \textcolor{blue}{\caption{List of Notations.}}
 \vspace{-5mm}
\begin{center}
\begin{tabular}{p{1.2cm}|p{6.2cm}||p{1.2cm}|p{6.3cm}}
\toprule %
\textbf{Notation} & \textbf{Definition} & \textbf{Notation} & \textbf{Definition} \\ 
[.3\normalbaselineskip]\hline
$K$& Number of all accessible clients&$E^{cp}_k$& Total energy consumption of client $k$ \\
\hline
$\mathcal{K}$ & Number of selected clients&$p_k^{cp}$ & Computing power of client $k$\\
\hline
$\mathcal{J}_k, \mathbf{x}, {y}$ & Local dataset of client $k$, input vector, output &$t_k^{cp}$ & Computing time of client $k$ \\
\hline
$\mathbf{w}$& Global FL model&$p_k^{bc}$ &Uploading power of client $k$ \\
\hline
$f$ & Loss function of each client&$\mathcal{N}_0$ & Noise power\\
\hline
$F_k$ & Total loss function of client $k$&$t_k^{up}$ & Uploading time of client $k$\\
\hline
$F$ & Total FL loss function&$E_k^{up}$ & Energy consumption of client $k$ for uploading\\
\hline
$J_k$ & Data samples of client $k$&$b_k$ & Allocated bandwidth to client $k$\\
\hline
$\mathbf{z}$ & The bias between global model and local update&${B}$ & Total network bandwidth\\
\hline
$i$ & Number of local iterations&$r_k$ & Uplink data rate of client $k$\\
\hline
$n$ & Number of communication rounds&$r_H$ & Downlink data rate\\
\hline $s$ & Constant size of the local update &$L_H$ & Computation density of HAPS server \\
\hline
$\eta$& Target local accuracy&$p_H^{bc}$ & Broadcasting power of HAPS server\\
\hline
$\epsilon$ &Target global accuracy&$t_H^{bc}$ & Broadcasting time of HAPS server\\
\hline
$v, M, u$ & Positive constant values&$a_k$ & Client selection binary variable\\
\hline
$\delta$ & Step size of SGD&$E_H$ & Broadcasting energy of HAPS server\\
\hline
$C_k$ & Number
of CPU cycles of client $k$&$p_H^{cp}$ & Computing power of HAPS server \\
\hline
$h_k$ & Channel power gain of client $k$&$t_H^{cp}$ & Computing time of HAPS server\\
\hline
$f_k$ & CPU capability of client $k$&${{E}}^{\max}_k$ & Maximum available energy of client $k$\\
\hline
$d_k$ & Distance between client $k$ and HAPS&${{E}}_H^{\max}$& Maximum available energy of HAPS server\\
\hline
$g_k$ & Rician channel power gain&$f^{\min},f^{\max}$ & Min. and max. computation capability of the clients \\
\hline
$\hat{h}_0$ & Reference free-space channel power gain&$\tau^{FL}$ & End-to-end FL delay (at a given round)\\
\hline
$\Delta d$ & Gaussian distributed random displacement of HAPS&$\tau^{DL}$ & Downlink delay\\
\hline
$d_0$ & Reference distance &$\tau^{UL}$ & Uplink delay\\
\hline
$f_H$ & Computing capability of HAPS server &$x_0$ & A real value associated with Taylor expansion\\
\hline$\zeta_k$ & Coefficient related to the hardware of client $k$&$\theta, \boldsymbol{\gamma}, \boldsymbol{\lambda}, \psi, \omega$ & Lagrangian multipliers\\
\hline$\zeta_H$ & Coefficient related to the hardware of HAPS server&$l_k$&Path loss model of client $k$\\
[.3\normalbaselineskip]
\bottomrule 
\end{tabular}
\end{center}
\vspace{-7mm}
\end{table*}
\vspace{-4mm}
\section{System Model}\label{sys}
\subsection{Network Topology Model}
\begin{figure}[!t]
    \centering
    \includegraphics[width=106mm,trim={2cm 1cm 0 0},clip]{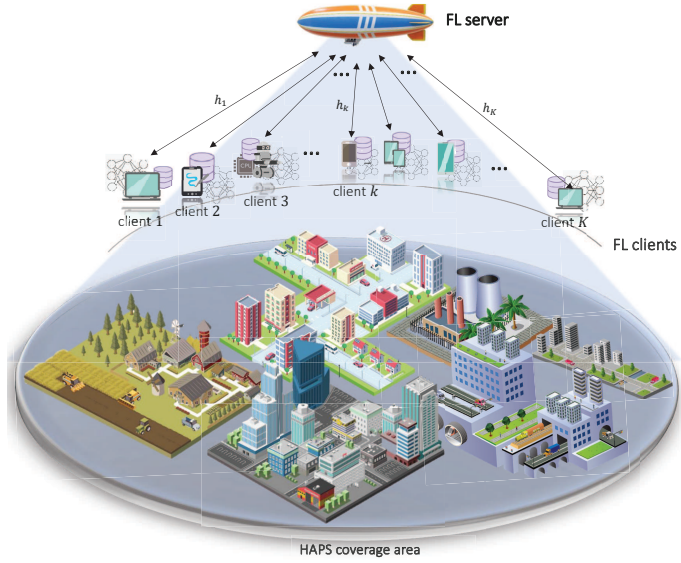} 
    \caption{The system model.}
    \vspace{-5mm}
    \label{fig1}
\end{figure}

We consider an FLSTRA system implemented on an integrated terrestrial-aerial heterogeneous\footnote{By heterogeneous, we mean that the devices in the network can have different computing or communication resources and data distribution.} network as illustrated in Fig.~\ref{fig1}. The integrated network consists of an FL server co-located with the HAPS in the stratosphere and $K$ single-antenna terrestrial devices randomly distributed on the ground within a wide coverage area of HAPS. The HAPS is responsible for operations, such as computing, data aggregation and broadcasting, and flying. Whereas, terrestrial devices are responsible for computing and communicating the trained local update to the FL server at the HAPS. In the FLSTRA system, a subset of these devices denoted by $\mathcal{K}$ such that $|\mathcal{K}|\leq K$, is selected as FL clients to connect with the HAPS server, and where $|\cdot|$ is the cardinality operator. Each client $k$ has a local dataset $\mathcal{J}_k$, such that  $|\mathcal{J}_k|=J_k$ data samples. Dataset $\mathcal{J}_k = \{{\bf{x}}_{kl}, {{y}}_{kl}\}^{J_k}_{l=1}$, where $({\bf{x}}_{kl},{{y}}_{kl}) \in \mathbb{R}^q \times \mathbb{R}$ represents the pair of input vector of size $q$ and scalar output corresponding to client $k$. It is important to note that our proposed framework is applicable to any HAPS, but we focus on the stratospheric platform in this paper to illustrate its unique features and advantages, such as large coverage areas, stable and long-duration flight, and line-of-sight communication channels. Nevertheless, the framework is easily extendable to other types of HAPS with varying altitudes and coverage areas. The analysis and solutions presented in this paper are independent of the specific type of HAPS used, making the framework adaptable to different HAPS based on application requirements. % \sout{, with $d$ the vectors' size. Moreover, each client is equipped with storage and computing units to support local observations, model training tasks, and communicate with the HAPS server.}
\vspace{-4mm}
\subsection{Federated Learning Model}
Adopting the traditional FL model \cite{trad-fl}, we define a vector $\mathbf{w}$ of size $q$ to collect the global training model or parameters of FLSTRA and introduce the loss function $f(\mathbf{w}, \mathbf{x}_{kl}, {y}_{kl})$ to access the performance of the learning taking place at client $k$.
%\sout{assess FL's performance over client $k$'s input and output vectors $\mathbf{x}_{kl}$ and $\mathbf{y}_{kl}$}.
We assume that all clients receive the same global model $\mathbf{w}$ and use the same loss function $f$. For different learning tasks, the loss function may have different definitions. For instance, $f(\mathbf{w}, \mathbf{x}_{kl}, {y}_{kl})=\frac{1}{2}(\mathbf{x}^T_{kl}\mathbf{w}-{y}_{kl})^2$ for linear regression and $f(\mathbf{w}, \mathbf{x}_{kl}, {y}_{kl})=-\log(1 + \exp(-{y}_{kl}\mathbf{x}^T_{kl}\mathbf{w}))$ for logistic regression. Since the dataset of client $k$ has $J_k$ data samples, the total loss function of client $k$ can be defined as
\vspace{-1mm}
\begin{align}
    F_k(\mathbf{w})= \frac{1}{J_k}\sum_{l=1}^{J_k}f(\mathbf{w},\mathbf{x}_{kl},{y}_{kl}).
\end{align}
%Note that function $f(\mathbf{w}, \mathbf{x}_{kl}, \mathbf{y}_{kl})$ is the loss function of client $k$ with one data sample and function $F_k(\mathbf{w}, \mathbf{x}_{k1}, \mathbf{y}_{k1}, \dots , \mathbf{x}_{kD_k} , \mathbf{y}_{kD_k} )$ is the total loss function of client $k$ with the whole local dataset. 
%For the sake of simplicity, we denote by $F_k(\mathbf{w})=F_k(\mathbf{w}, \mathbf{x}_{k1}, \mathbf{y}_{k1}, \dots , \mathbf{x}_{kD_k} , \mathbf{y}_{kD_k})$.
%is denoted by $F_k(\mathbf{w})$ for simplicity of notation.\\

To deploy FL-based training, it is necessary to train its global model $\mathbf{w}$ by inviting a set of clients to participate in the training process. Training is performed via all invited clients without sharing their raw data to generate a new common global model. Accordingly, the FLSTRA learning problem can be formulated as
\vspace{-1mm}
\begin{align}\label{global}
    \operatorname*{min}_{\mathbf{w}\in{\mathbb{R}^{q}}} \Big\{F(\mathbf{w})\overset{\Delta}{=}\sum_{k\in \mathcal{K}}\frac{J_k}{J}F_k(\mathbf{w})=\frac{1}{J}\sum_{k\in \mathcal{K}}\sum_{l=1}^{J_k}f(\mathbf{w},\mathbf{x}_{kl},{y}_{kl})\Big\},
\end{align}
where $J =\sum_{k=1}^K J_k= |\cup_{k\in \mathcal{K}}\mathcal{J}_k|$ is the total number of data samples of all involved or selected clients. 

To solve problem \eqref{global}, we develop the communication and computation resource-aware (CCRA-FL) algorithm by adapting the federating averaging (FedAvg)~\cite{intro7} algorithm. FedAvg consists of alternating between a number of local gradient updates at clients, followed by a model averaging update at the server. Specifically, CCRA-FL uses stochastic gradient descent (SGD) as the optimizer to generate local updates on each client. One disadvantage of batch gradient descent is that it performs redundant computations for large datasets by recomputing gradients for similar examples before each parameter update. SGD, on the other hand, eliminates this redundancy by performing one update at a time~\cite{intro7-b}. This makes SGD usually much faster than gradient descent and more suitable for large-scale datasets where computational efficiency is crucial. Further, in the CCRA-FL algorithm, the FedAvg parameters are optimized by solving the network resource optimization problem, which is developed in Section III, subject to communication and computation constraints. %, and can also be used to learn online \textcolor{red}{(what does it mean learning online? This is confusing as you never talked about learning offline or online previously)}. 

CCRA-FL algorithm solves \eqref{global} in an iterative manner until the desired global target accuracy of $\epsilon > 0$ for the global model is achieved. We denote the total number of iterations (i.e., the number of global FL iterations or communication rounds) required for convergence by $n$. In particular, at communication round $n$:
\begin{enumerate}
    \item Each client $k \in \mathcal{K}$ performs SGD locally to obtain its local update. 
    \item All the selected clients $|\mathcal{K}|$ under the coverage of HAPS upload their updated model parameters using the frequency-division medium access (FDMA) scheme to the FL server at HAPS. 
    \item Upon receiving all the local updates, the FL server at HAPS aggregates them and forms new updated global model parameters $\mathbf{w}^{(n)}$ corresponding to communication round $n$.
    \item The HAPS server broadcasts the updated global model parameters back to all the clients.
    \item Each client solves a local optimization problem with a given local target accuracy $\eta$ at each communication round. 
\end{enumerate}
In the rest of the paper, we refer to FL global iteration as the communication round and the local FL iteration as the local iteration.

We further assume that the global model has the same size as the local update and is constant during all communication rounds, i.e., $|\mathbf{w}^{(n)}|=s$ bits, $\forall n$. Each client $k \in \mathcal{K}$ solves the following local optimization problem during each communication round:
\vspace{-1mm}
\begin{align} \label{local}
    \min_{\mathbf{z}_k^{(n)}\in{\mathbb{R}^{q}}} \Big\{G_k(\mathbf{w}^{(n)},\mathbf{z}_k^{(n)})\delequal F_k(\mathbf{w}^{(n)}+\mathbf{z}_k^{(n)}) -(\nabla F_k(\mathbf{w}^{(n)})-\xi\nabla F(\mathbf{w}^{(n)}))^T\mathbf{z}_k^{(n)}\Big\}, 
\end{align}
where $\xi$ is a constant, and $\mathbf{z}_k^{(n)}$ is the bias term between the global model and local update of client $k$ at communication round $n$. Precisely, $\mathbf{w}^{(n)} + \mathbf{z}_k^{(n)}$ is the local update of client $k$ at communication round $n$. $\nabla F_k(\mathbf{w}^{(n)})$ is the gradient of function $F_k(\mathbf{w}^{(n)})$ at point $\mathbf{w}^{(n)}$. Each client employs the SGD method to solve \eqref{local}. To analyze the convergence rate of the SGD method, we impose the following standard assumptions on the loss function $F_k(.)$:
\vspace{-3mm}
\begin{assmp}[\cite{assumps}]\label{assmp1}
$F_k(\cdot)$ is $M$-Lipschitz continuous and $u$-strongly convex if
\begin{align}
    &F_k(\mathbf{w})\leq F_k(\mathbf{w}^\prime) + \langle \nabla F_k(\mathbf{w}^\prime),\mathbf{w}-\mathbf{w}^\prime\rangle + \frac{M}{2}\|\mathbf{w}-\mathbf{w}^\prime\|^2, \label{assm1}\\
    & F_k(\mathbf{w})\geq F_k(\mathbf{w}^\prime) + \langle \nabla F_k(\mathbf{w}^\prime),\mathbf{w}-\mathbf{w}^\prime\rangle + \frac{u}{2}\|\mathbf{w}-\mathbf{w}^\prime\|^2, \label{assm2}
\end{align}
where the Lipschitz constant $M$ and $u$ are positive constants, 
%$\mathbf{w}$ refers to as the model parameter,
$\langle \cdot,\cdot \rangle$ is the inner product operation, $\|\cdot \|$ denotes the Euclidean norm, and $\nabla F_k(\mathbf{w}^\prime)$ is the gradient of $F_k(\cdot)$ with respect to $\mathbf{w}^\prime$.
\end{assmp}

According to Assumption~\ref{assmp1}, problem~\eqref{local} can be solved with a given target local accuracy of $\eta$ (i.e., the lower the value of $\eta$, the higher the local accuracy and vice-versa) if the maximum number of SGD iterations $i$ is bounded from below as \cite{fl1}
\vspace{-3mm}
\begin{align}\label{fl-bound1}
    i\geq v\log_2\Big(\frac{1}{\eta}\Big),
\end{align}
where $v=\frac{2}{(2-M\delta)\delta u}$, $\delta<\frac{2}{M}$ is the step size in the SGD method, and $\eta$ 
satisfies the following convergence condition for the SGD method in the CCRA-FL algorithm:
\vspace{-2mm}
\begin{align}\label{eta-bound}
    G_k(\mathbf{w}^{(n)},\mathbf{z}_k^{(n,i)}) - G_k(\mathbf{w}^{(n)},\mathbf{z}_k^{(n)*}) \leq \eta \Big[ G_k(\mathbf{w}^{(n)},\mathbf{z}_k^{(n,0)}) -G_k(\mathbf{w}^{(n)},\mathbf{z}_k^{(n)*})\Big],
\end{align}
where $\mathbf{z}_k^{(n)*}$ is the optimal solution of local optimization problem~\eqref{local}. Further details about the proof of aforementioned bounds~\eqref{fl-bound1} and~\eqref{eta-bound} can be found in~\cite{fl1}. \eqref{fl-bound1} implies that to achieve a highly accurate local update (i.e., lower value of $\eta$), the number of local iterations has to be higher. 

Under the assumptions mentioned above, we present an upper bound on the convergence rate of the CCRA-FL algorithm in Theorem 1. Specifically, Theorem 1 provides an upper bound on the difference between the final loss of the CCRA-FL algorithm and the optimal global model. This difference represents an upper bound on the convergence rate of the FL system and is a commonly used metric for measuring the performance of such systems~\cite{conv-rate}. It is worth noting that the loss function for the optimal global model is always less than or equal to the loss function for any global model since the optimal global model is the one that minimizes the global loss function. Hence, the difference between the final and optimal loss in Theorem 1 is always positive, representing the upper bound on the convergence rate of the CCRA-FL algorithm.
%The convergence rate is defined as the number of global iterations needed for the algorithm~\ref{alg1} in solving problem \eqref{global}. According to Theorem 1 in~\cite{fl1}, the number of global iteration increases with the target local accuracy $\eta$ at the rate of $\frac{1}{(1 - \eta)}$ such that it is possible to derive the following lower bound for the number of global iterations
%\begin{align}\label{fl-bound2}
  %  N\geq \frac{\alpha}{1-\eta},
%\end{align}
%where $\alpha=\frac{2L^2}{y^2\zeta}\ln \frac{1}{\epsilon}$. According to this Theorem, the FL performance depends on parameters $L,\: y ,\: \zeta ,\: \epsilon$ and $\eta$, and it provides a general convergence rate for FL with an arbitrary $\eta$.
\begin{thm}\label{conv-thm}
If we run CCRA-FL Algorithm for $n$ communication rounds, its convergence rate with $|\mathcal{K}^{(.)}|$ selected clients (under the coverage of HAPS) at each communication round, and given target local accuracy $\eta$, can be upper-bounded as
\vspace{-3mm}
\begin{align} \label{conv}
    F(&\mathbf{w}^{(n)})- F(\mathbf{w}^*)\leq \epsilon,
\end{align}
where
\vspace{-6mm}
\begin{align} \label{eps}
    &\epsilon = \Big(1-\frac{(1-\eta)u^2\xi}{2M^2}\Big)^{n}\Big[F(\mathbf{w}^{(0)}) - F(\mathbf{w}^*)\Big]+\frac{M\xi-u}{2\xi}\nonumber\\ &\times\Bigg[\sum_{n^\prime=0}^{n-1}\frac{1}{|\mathcal{K}^{(n^\prime)}|}\Big(1-\frac{(1-\eta)u^2\xi}{2M^2}\Big)^{n-n^\prime-1}\sum_{k\in \mathcal{K}^{(n^\prime)}}\|\mathbf{z}_k^{(n^\prime)}\|^2\Bigg],
\end{align}
is the target convergence upper bound of the FL process under the condition $u\leq M\min\Big\{\zeta,\sqrt{\frac{2}{\zeta}}\Big\}$, which guarantees $\epsilon\geq 0$. $\mathbf{z}_k^{(n^\prime)}$ is the bias term between the global model and local update of client $k$ at communication round $n^\prime$, $n^\prime\in\{0,\dots,n\}$. 
\end{thm}
\vspace{-5mm}
{\proof See Appendix~\ref{app:1a}.}

We observe that Theorem~\ref{conv-thm} includes a new term  on the right-hand side of \eqref{eps} that is not present in the result presented in \cite{fl1}. This additional term captures the effect of the number of selected clients on the convergence rate of FL in HAPS-based systems. Our analysis indicates that an increase in the number of selected clients $|\mathcal{K}^{(n)}|$ during each communication round results in the disappearance of the biased term, leading to a faster convergence rate. However, this improvement comes at the cost of increased utilization of network resources, higher delay, and more energy consumption. It is important to note that this dependency on the number of selected clients is unique to HAPS-based FL, as it is a result of its large footprint, providing the luxury of choosing different numbers of participants for each round of communication.
\begin{algorithm}[!t] 
 \caption{CCRA-FL Algorithm.}
 \label{alg1}
 \begin{algorithmic}[1]
  \STATE Initialization $\rightarrow$ Global model $\mathbf{w}^{(0)}$ and $n=0$. \\
{\bf{\textit{LOOP Process}}}
\STATE Optimization $\rightarrow$ HAPS finds $\eta^{(n)}$, $a_k^{(n)}$, and resource allocation policies by using Algorithm~\ref{alg2}.\\
\STATE Broadcasting $\rightarrow$ HAPS transmits $\mathbf{w}^{(n)}$ to the selected clients.\\
\STATE Local computing $\rightarrow$ Each selected client $k\in\mathcal{K}^{(n)},\: |\mathcal{K}^{(n)}|=\sum_{k=1}^K a_k^{(n)}$, calculates $\nabla F_k(\mathbf{w}^{(n)})$ and uploads it to HAPS server.
  \STATE Aggregation $\rightarrow$ HAPS server aggregates all the gradients and calculates $\nabla F(\mathbf{w}^{(n)})=\frac{1}{|\mathcal{K}^{(n)}|}\sum_{k\in\mathcal{K}^{(n)}}\nabla F_k(\mathbf{w}^{(n)})$.\\
  {\bf{\textit{Parallel For}}} $k\in\mathcal{K}^{(n)}$\\
  \STATE \quad Initialization $\rightarrow$  $i=0$ and $\mathbf{z}_k^{(n,0)}=0$.\\
  \quad{\bf{\textit{LOOP Process}}} (SGD Algorithm)
  \STATE \qquad Update $\rightarrow$ $\mathbf{z}_k^{(n,i+1)}=\mathbf{z}_k^{(n,i)}-\delta\nabla G_k(\mathbf{w}^{(n)},\mathbf{z}_k^{(n,i)})$ and $i=i+1$.
  \STATE \quad \textbf{Until}  $i= v\log_2(\frac{1}{\eta^{(n)}})$, i.e., local accuracy $\eta^{(n)}$ is achieved.
  \STATE \quad Local update $\rightarrow$ Denote  $\mathbf{z}_k^{(n)}=\mathbf{z}_k^{(n,i)}$ and upload $\mathbf{z}_k^{(n)}$ to HAPS server.\\
  {\bf{\textit{End For}}}
  \STATE New global model $\rightarrow$  $\mathbf{w}^{(n+1)}=\mathbf{w}^{(n)}+\frac{1}{|\mathcal{K}^{(n)}|}\sum_{k\in\mathcal{K}^{(n)}}\mathbf{z}_k^{(n)}$.
  \STATE set $n=n+1$.
  \STATE \textbf{Until} The target convergence upper bound $\epsilon$ achieved in~\eqref{conv}.
  %the global accuracy $\epsilon_0$ of problem~\eqref{global} is obtained.
 \end{algorithmic}
 \end{algorithm}
%%%%%%%%%%%%%%
\vspace{-5mm}
\subsection{Computation Model}
%\textcolor{red}{Pelase add 1 or 2 sentences to explain that different computing powers are in clients and server...}
The FLSTRA process includes local gradient computation at clients and local updates aggregation by the FL server at the HAPS. Therefore, both the HAPS server and clients perform computational tasks. We describe their computation models in the following.
\subsubsection{Computation model at clients}
%{\bf{\textit{Clients.}}} 
For the local computation of client $k$, we denote $f_k$ (in CPU cycles per second) as the CPU computing capability, 
%$D_k$ as the number of data samples, 
and $C_k$ as the number of CPU cycles needed to process one data sample. At any communication round, the computation time for processing $J_k$ data samples with $i$ local iterations is
\vspace{-3mm}
\begin{align}
    t^{cp}_k=\frac{iC_kJ_k}{f_k}, \: \forall k.
\end{align}
%\textcolor{red}{The presence of the term $J_k$ is confusing here, as it says in each round, the WHOLE local dataset is trained, which is wrong! Please clarify what do you want to say here!} \textcolor{blue}{ Answer: Following similar works, the whole data sample is trained at each client.}
The corresponding energy consumption at client $k$, in a communication round, can be calculated as follows \cite[Lemma 1]{comp}:
\vspace{-4mm}
\begin{align}
    E^{cp}_k=t^{cp}_kp^{cp}_k, \:\forall k,
\end{align}
where $p^{cp}_k=\zeta_kf^3_k$ is the computing power \footnote{According to \cite[Lemma 1]{comp}, the energy consumption of each client at each communication round can be calculated as $E_k^{cp}=i\zeta_kC_kJ_kf_k^2, \: \forall k$. Thus, the computing power consumption can be determined as $p_k^{cp}=E_k^{cp}/t_k^{cp}$, such that $p^{cp}_k=\zeta_kf^3_k, \: \forall k.$} of client $k$, and $\zeta_k$ is a coefficient depending on the hardware and chip architecture of client $k$.
\vspace{-1mm}
%\\ \indent {\bf{\textit{HAPS server:}}} 
\subsubsection{Computation model of FL server at HAPS}

At each communication round, the local updates which HAPS receives are decoded and aggregated to generate a new global model. The processing time of this task can be calculated as
\vspace{-3mm}
\begin{align}
    t^{cp}_H=\frac{L_HQ_H}{f_H},
\end{align}
where $Q_H=s\sum_{k=1}^Ka_k$ is the total number of bits needed to be processed at HAPS at a given communication round with $s$ as the size (in bits) of the local update of a client received by the HAPS. We use the binary variable $a_k$ to indicate whether or not client $k$ is selected for FL training and to upload its local update to the HAPS server during a given communication round. $L_H$ is the computation density (in CPU cycle/bit)~\cite{comp density}, and $f_H$ is the CPU computing capability of the FL server at HAPS~\cite{comp-dens2}.  %, then $a_k=1$, otherwise $a_k=0$.
\vspace{-6mm}
\subsection{Channel and Communication Models}
We consider both large-scale and small-scale fading effects to model the communication channels between terrestrial clients and HAPS. Let $d_k$ denote the distance between client $k$ and HAPS, where $k\in \{1,\dots, K\}$.
%such that $\mathcal{K}=\{k|a_k\neq 0\}$ and $|\mathcal{K}|=\sum_{k=1}^K a_k$.
Accordingly, the corresponding channel power gain at a given global round can be calculated as
\vspace{-1mm}
\begin{align}
    h_k=g_k\hat{h}_0\Big(\frac{d^\prime_k}{d_0}\Big)^{-2}, \: \forall k,
\end{align}
where $d^\prime_k=d_k+\Delta d$ is the effective distance between client $k$ and HAPS with $\Delta d$ as the random displacement of HAPS from its original position due to the impact of natural phenomenons, such as  stratospheric winds~\cite{disp-haps}. We assume that $\Delta d$ follows a Gaussian distribution with zero-mean and variance $\sigma^2$. $\hat{h}_0$ denotes the reference free-space channel power gain at $d_0 = 1$ km. %$\sigma^2$ is basically related to the speed of HAPS displacement during iteration $n$ and defined as follows:
%\begin{align*}
 %   \sigma=
%\end{align*}
$g_k$ captures the small-scale fading effects and follows the Rician block fading model. We adopt a block fading channel model where channel gains are constant within a communication round but vary independently from round to round. The channel variation across rounds is characterized by the Rician model, which captures small-scale fading, and the average path loss, which accounts for large-scale fading~\cite{Rician1}. This model effectively accommodates the changes in the channel conditions across rounds and enables us to evaluate the performance of our system under realistic conditions~\cite{Rician2}. 
%We assume that the channel power gains are independent of communication rounds and constant within a round. % and the corresponding probability density function (PDF) at each iteration can be calculated as
%\begin{align}
 %   f_{g_k}=\frac{(\mathbb{K}+1)\exp^{-\mathbb{%K}}}{\lambda_k}\exp^{-\frac{(\mathbb{K+1})}{}}
%\end{align}

%\\ \indent {\bf{\textit{Uplink.}}} 

\subsubsection{Uplink Transmission} To mitigate the delay caused by clients taking turns to upload their updates in time-division medium access (TDMA), we adopt FDMA for the uplink transmission in our system model. With FDMA, after local computation, multiple clients can simultaneously upload their local updates to the HAPS server on separate frequency bands for aggregation, reducing the overall delay and improving the efficiency of the system~\cite{FDMA}. Accordingly, the instantaneous achievable uplink rate of client $k$ is given as
\vspace{-1mm}
\begin{align}
    r_k=b_k\log_2\Big(1+\frac{h_kp_k^{up}}{b_k\mathcal{N}_0}\Big), \: \forall k,
\end{align}
where $b_k$ is the allocated bandwidth to client $k$, $p_k^{up}$ is the transmit power, and $\mathcal{N}_0$ is the noise power spectral density in dBm/Hz. Further, the following constraint must hold to ensure that each selected client uploads its local update within the uploading time $t^{up}_k$:
\vspace{-1mm}
\begin{align}
     r_kt^{up}_k\geq s, \: \forall k.
\end{align}
Accordingly, the corresponding energy consumed by client $k$ for uploading its local update can be expressed as 
\vspace{-7mm}
\begin{align}
    E^{up}_k=t^{up}_kp_k^{up}, \: \forall k.
\end{align}
%\indent {\bf{\textit{Downlink.}}} 
\subsubsection{Downlink Transmission}
After aggregating all the local updates and generating a new global model, the HAPS server broadcasts the new model to all clients. The HAPS server can adapt its data rate by adjusting its broadcasting power according to the channel quality of the selected clients. This is achieved by adapting the data rate to the client with the worst instantaneous signal-to-noise ratio (SNR). Hence, the corresponding instantaneous achievable downlink rate at each communication or global FL round can be determined as 
\vspace{-3mm}
\begin{align}
    r_H=\min_{\substack{k\in\mathcal{K} }}~{{B}}\log_2\Big(1+\frac{p_H^{bc}h_k}{{{B}}\mathcal{N}_0}\Big),
\end{align}
where $p_H^{bc}$ is the power allocated for broadcasting the global update, and ${{B}}$ is the total available network bandwidth. The downlink rate equation ensures that the HAPS server always broadcasts at the highest possible data rate that guarantees connectivity with all the selected clients under its coverage. Therefore, to ensure the transmission of new global model $\mathbf{w}^{(n)}$ to all selected clients within the broadcasting time $t_H^{bc}$, the following constraint should hold:
\vspace{-3mm}
\begin{align}
    r_Ht_H^{bc}\geq s,
\end{align} 
where $s$ is the size of the global update at a given communication round. Accordingly, the energy consumption of the HAPS can be given by
\vspace{-3mm}
\begin{align}
    E_H=p_H^{cp}t_H^{cp}+p_H^{bc}t_H^{bc},
\end{align}
where $p_H^{cp}=\zeta_H f_H^3$ is the HAPS's computing power consumption to process the received local updates and generate a new global model, with $\zeta_H$ a coefficient depending on the hardware and chip architecture of the HAPS server~\cite{intro-haps3}. Since we assume that flying power is fixed over all communication rounds~\cite{fixed-fly}, it can be ignored in the HAPS total energy consumption.
%we only consider communication and computation power consumption.
\vspace{-3mm}
\section{Delay Minimization Problem Formulation and Proposed Solution}\label{prob}
According to Theorem~\ref{conv-thm} and the intrinsic large footprint of HAPS, a massive number of accessible clients can be invited to the FLSTRA process through single-hop communications, thus leading to a higher FL convergence rate. Moreover, Theorem~\ref{conv-thm} indicates a trade-off between the communication (and computation) efficiency and the FL convergence rate. Hence, in this work, we aim to improve the FL convergence rate by optimizing the number of selected clients and optimizing the resources while minimizing the end-to-end FL delay at every communication round. This delay includes both communication and computation times at both uplink and downlink at each communication round. In this section, we first formulate the end-to-end delay minimization problem for FLSTRA. Secondly, due to the non-convexity of the corresponding problem, we develop an iterative algorithm to solve it.
\vspace{-5mm}
\subsection{Problem Formulation}
We aim here to develop a joint resource allocation and client selection strategy that minimizes the end-to-end FL delay in a communication round while achieving the target global accuracy $\epsilon$. Let $\tau^{FL}$ be the end-to-end delay of the FL process at a given communication round, which is defined as {\small{$\tau^{FL}=\tau^{UL}+\tau^{DL}$}}. $\tau^{UL}$ and $\tau^{DL}$ are the uplink and downlink delays, respectively, and are determined as follows:
\vspace{-4mm}
\begin{align}
    &\tau^{UL}=\max_k~ a_k(t_{k}^{up}+t^{cp}_k),\\
    &\tau^{DL}= t_H^{bc}+t_H^{cp},
\end{align}
where $t_H^{cp}$ and $t_H^{bc}$, respectively, are the time limits allocated by the HAPS for computing, i.e., aggregating the received local updates and broadcasting the global model parameters $\mathbf{w}^{(n)}$ to participating clients at the given communication round $n$.
%\\
%\indent Moreover, since the FL algorithm involves the accuracy of local computation and the result aggregation, it is hard to calculate the exact number of iterations needed for convergence. However, from delay minimization perspective, it is always efficient to minimize the number of iterations. Thus,
%using \eqref{fl-bound1}, the number of local iterations $I_k[n]$ of client $k$ at iteration $n$ can be determined as
%\begin{align}
 %   I_k[n]=v\log_2(\frac{1}{\eta[n]}), \: \forall k.
%\end{align}
%Similarly, by using~\eqref{fl-bound2}, to minimize the required number of global iterations, we consider the most delay-resulting scenario, i.e., $\eta=\eta^{\max}$, where $0<\eta^{\max}<1$, is the maximum desired target accuracy. Therefore,  the required number of total global iterations $N$ can be calculated as 
%\begin{align}\label{approx}
%    N\leq \frac{\alpha}{1-\eta}.
%\end{align}
%\indent 

Now, we formulate the end-to-end delay minimization problem at a given communication round, considering both uplink and downlink transmissions, as follows:
\vspace{-4mm}
\begin{subequations}
\begin{align} \label{p1-main}
    \operatorname*{min}_{p^{bc}_H,t^{bc}_H, f_H,\eta,\mathbf{p},\mathbf{f},\mathbf{b},\mathbf{a},\mathbf{t}}\quad &  \tau^{FL}\\
     \text{s.t.}\quad 
     %& \frac{a_k[n]\alpha}{1-\eta}(\frac{vC_kD_k\log_2(\frac{1}{\eta})}{f_k[n]}+t_k^{up}[n])+t_H^{cp}+t_H^{tr}\leq T ,\label{p1-a1}\\
    &a_k\Big[t_k^{up}b_k\log_2(1+\frac{p_k^{up}h_k}{b_k\mathcal{N}_0})-s\Big]\geq 0, \forall k,\label{p1-const1}\\
        & a_k\Big[t_H^{bc} B\log_2(1+\frac{p_H^{bc}h_k}{B \mathcal{N}_0})-s\Big]\geq 0, \: \forall k,\label{p1-const2} \\
    &a_k\Big[t^{up}_kp_k^{up}+p_k^{cp}t_k^{cp}\Big]\leq E_k^{\max}, \: \forall k, \label{p1-const3}\\
        & p_H^{cp}t^{cp}_H+p_H^{bc}t_{H}^{bc}\leq E_H^{\max},\label{p1-const4}\\
         & \frac{vC_kJ_k\log_2(\frac{1}{\eta})}{f_k}\leq t_k^{cp},\: \forall k,\label{p1-const5}\\
    & \frac{L_HQ_H}{f_H}\leq t^{cp}_H, \label{p1-const6}\\
    %&0\leq P\leq P^{\max}_{H},\\
    & \sum_{k=1}^{K}a_kb_k\leq B, \label{p1-const7}\\
    % & N\geq \frac{\alpha}{1-\eta}, \label{p1-N}
      % &\: 0\leq N\leq N^{\max},\\
     &  a_k\in\{0,1\}, \: \forall k,\label{p1-const8}\\
     &0\leq p_k^{up}\leq p^{\max}_k, \: \forall k,\label{p1-const9}\\
     %&0\leq p_H^{bc}\leq p^{\max}_k, \: \forall k,\\
    & f^{\min}_k\leq f_k\leq f^{\max}_k, \: \forall k,\label{p1-const10}\\
     & 0<\eta\leq 1,\: \forall k, \label{p1-const11}\\
     & t_k^{up},t_H^{bc}\geq 0,\: b_k\geq 0, \: \forall k,
\end{align}
\label{Pr20}
\end{subequations}
where $\mathbf{b}=[b_1,\dots, b_K]$, $\mathbf{f}=[f_1,\dots, f_K]$,
 $\mathbf{a}=[a_1,\dots, a_K]$,
$\mathbf{t}=[t_1^{up},\dots, t^{up}_K]$, and $\mathbf{p}=[p^{up}_1,\dots, p^{up}_K]$. \eqref{p1-const1} and \eqref{p1-const2} are the uplink and downlink data transmission constraints, respectively. Constraint \eqref{p1-const3} ensures the local energy consumption constraint at each selected client is not exceeding the maximum available energy (i.e., $E_k^{\max}$) at client $k$ in each communication round. Constraint \eqref{p1-const4} ensures the total energy consumption, per communication round, at HAPS is not exceeding $E_H^{\rm max}$. \eqref{p1-const5} and \eqref{p1-const6} represent the constraints on the computation times at the clients and HAPS server, respectively. Constraint \eqref{p1-const7} ensures that the bandwidth distribution among clients is not more than $B$ Hz. Constraints \eqref{p1-const8} restrict the client selection decision variable to take only binary values, while constraints \eqref{p1-const9}-\eqref{p1-const10} represent the average transmit power limits and the maximum local computation capacity of clients, respectively. $p_k^{\max}$ is the maximum power available at client $k$ for uploading. Finally, the target local accuracy constraint is given by \eqref{p1-const11}.
\vspace{-3mm}
%%%%%%%%%%%%%%%
 \begin{algorithm}[!t]  
 \caption{Proposed Iterative Algorithm.}
 \label{alg2}
 \begin{algorithmic}[1]
  \STATE Initialization $\rightarrow$ HAPS parameters $(p_H^{bc^{(0)}},t_H^{bc^{(0)}},f_H^{(0)})$, resource allocation parameters $(\mathbf{f}^{(0)},\mathbf{t}^{(0)}, \mathbf{p}^{(0)},\mathbf{b}^{(0)})$, user selection $\mathbf{a}^{(0)}$, target local accuracy $\eta^{(0)}$, $l^{\max}$, and $l=0$.\\
  {\bf{\textit{LOOP Process}}}
\STATE Solve \eqref{Pr21} $\rightarrow$ Obtain the optimal ($\eta^{(l+1)},\mathbf{t}^{(l+1)}$) for given $(p_H^{bc^{(l)}},t_H^{bc^{(l)}},f_H^{(l)}, \mathbf{t}^{(l)},\mathbf{p}^{(l)},\mathbf{b}^{(l)}, \mathbf{a}^{(l)})$. 
  \STATE Solve \eqref{af} $\rightarrow$ With given ($\eta^{(l+1)},\mathbf{t}^{(l+1)}$), find optimal ($\mathbf{f}^{(l+1)}, \mathbf{a}^{(l+1)}$).
  \STATE Solve \eqref{Pr27} $\rightarrow$ Find the optimal ($\mathbf{p}^{(l+1)}, \mathbf{b}^{(l+1)}$) for given $(\eta^{(l+1)},\mathbf{t}^{(l+1)},\mathbf{f}^{(l+1)}, \mathbf{a}^{(l+1)})$.
  \STATE Solve \eqref{Pr30} $\rightarrow$ Obtain the optimal HAPS parameters  $(p_H^{bc^{(l+1)}},t_H^{bc^{(l+1)}},f_H^{(l+1)})$ with given $(\eta^{(l+1)},\mathbf{t}^{(l+1)},\mathbf{f}^{(l+1)}, \mathbf{a}^{(l+1)},\mathbf{p}^{(l+1)},\mathbf{b}^{(l+1)})$.
  \STATE Set $l=l+1$.
  \STATE \textbf{Until} Convergence=\textbf{true} or $l=l^{\max}$
  \end{algorithmic}
 \end{algorithm}
%%%%%%%%%%%%%%%
\subsection{Proposed Iterative Algorithm}
We observe that problem \eqref{Pr20} is a mixed integer non-linear program (MINLP), which cannot be solved optimally with rapid convergence through standard convex optimization schemes. We propose an efficient iterative algorithm wherein we alternately optimize four sub-problems until the algorithm converges. In the following, we discuss the steps involved in solving the four sub-problems. 
\begin{enumerate}[leftmargin=*]
% uses Branch and Bound mathematical methods (BB) to give a optimal solution, however, the complexity of BB is equal to exhaustive search, which can not be used in real network. 
%The objective of this paper is to minimize the the total delay via designing an intelligent FL resource allocation and user selection algorithms along with computation algorithm. This enables HAPS to make network management decisions at each FL global round considering its power limitations and jitter effect, and clients' local features such as computational capability.\\
\item{\textit{Optimal client uploading time and local accuracy $(\mathbf{t}^*,\eta^*)$}}: In the first step, we fix the variables associated with HAPS (i.e., $\{p_H^{bc},t_H^{bc},f_H\}$), the resource allocation and user selection variables (i.e., $(\mathbf{p}, \mathbf{f}, \mathbf{b}, \mathbf{a}, \mathbf{t})$), and then jointly optimize clients uploading time and target local accuracy by solving the following first sub-problem:
\vspace{-4mm}
\begin{subequations}
\begin{align} \label{p2}
    \operatorname*{min}_{\mathbf{t},\eta}\quad  &  \Big[\operatorname*{max}_k~  a_k\Big(t_k^{up}+\frac{vC_kJ_k\log_2(\frac{1}{\eta})}{f_k}\Big)\Big]\\
     \text{s.t.\quad}&
    % & \frac{\alpha}{1-\eta}(\frac{vC_kD_k\log_2(\frac{1}{\eta})}{f_k[n]}+t_k^{up}[n])\leq \frac{E_H[n]}{P_H^{tr}[n]+P_H^{cp}[n]},\: \forall k,n.\label{a1}\\
    t_k^{up}\geq t_k^{\min},\: \forall k,\label{a2}\\
    &a_k\Big[t^{up}_kp_k^{up}+\zeta_kvC_kJ_k\log_2(\frac{1}{\eta})f^2_k\Big]\leq {{E}}_k^{\max},\label{p2-const2}\: \forall k,\\
     & 0<\eta\leq 1, \label{p2-const3}
\end{align}
\label{Pr21}
\end{subequations}
where
\vspace{-8mm}
\begin{align}
    t_k^{\min}=\frac{a_ks}{b_k\log_2(1+\frac{p_k^{up}h_k}{b_k\mathcal{N}_0})}, \: \forall k.
\end{align}
In problem (\ref{Pr21}), we aim to minimize $\tau^{UL}$, reflected in minimizing the uplink delay of the worst participating client. It can be observed that it is always efficient to allocate the minimum time for local uploading from a delay minimization perspective. Hence, the closed-form optimal solution of \eqref{p2} can be obtained using the following Theorem~\ref{Thm1}.
\begin{thm}\label{Thm1}
 The optimal solution $(\mathbf{t}^*=[t_k^{up^*}]_{1\times K},\eta^*)$ of problem \eqref{Pr21} satisfies
 \vspace{-1mm}
\begin{align}
    &t^{up^*}_k= t^{\min}_k, \: \forall k,
\end{align}
and
\vspace{-5mm}
\begin{align}\label{eta}
    &\eta^*=\frac{\beta_1+\beta_2}{\theta\ln{2}},
\end{align}
where $\beta_1=\operatorname*{max}_k \frac{a_kvC_kJ_k}{f_k}$ and $\beta_2=\sum_{k\in \mathcal{K}}\lambda_{k}a_k\zeta_kvC_kJ_kf_k^2$. Moreover, $\boldsymbol{\lambda}=[\lambda_k]_{1\times K}$ and $\theta$ are Lagrangian multipliers associated with constraints~\eqref{p2-const2} and \eqref{p2-const3}, respectively. 
\end{thm}
{\proof See Appendix~\ref{app:1}.}
\item{\textit{Optimal client selection and computing capability $(\mathbf{a}^*,\mathbf{f}^*)$}}:
%\indent Step 1:
%%   &\operatorname*{min}_{P,F,B,A} \sum_{n=1}^N \operatorname*{max}_k a_k[n](t_k^{up}[n]+\frac{vC_kD_k\log_2(\frac{1}{\eta})}{f_k[n]})\\
%%%    &a_k[n]\Big(t^{up}_k[n]p_k[n]+\zeta_kvC_kD_k\log_2(\frac{1}{\eta})f^2_k[n]\Big)\leq \mathbb{E}_k^{\max}, \: \forall k, n,\\
 %      & \sum_{k=1}^{K}a_k[n]b_k[n]\leq \mathbb{B}, \: \forall n, \label{a4}\\
%    & t_H^{tr}[n] \Big (\min_k a_k[n]\mathbb{B}\log_2(1+\frac{P_H^{tr}[n]h_k[n]}{\mathbb{B} N_0})\Big)\geq |w[n]|, \: \forall n, \\
    %&0\leq P\leq P^{\max}_{H},\\
   % & t^{cp}_k+t^{up}_k\leq T,\\ 
    %& P_H^{cp}[n]t^{cp}_{H}[n]+P_H^{tr}[n]t^{tr}_{H}[n]\leq E_H[n], \: \forall n,\label{a6}\\
    %&\sum_{n=1}^N E_H[n]\leq E^{\max},\label{a7}\\
%     &  a_k[n]\in\{0,1\}, \: \forall n,k,\label{a8b}
%      \\&0\leq p_k[n]\leq p^{\max}_k, \: \forall n,k,\label{a3}\\
%    & f^{\min}_k\leq f_k[n]\leq f^{\max}_k, \: \forall n,k,\label{a5}\\
 %   % & 0<\eta\leq 1, \label{a9}\\
%     & b_k[n]\geq 0, \: \forall n,k.
%\end{align}
%\indent 
In the second step, we aim to jointly determine the optimal client selection policy (i.e., $\mathbf{a}$), and local computing capability (i.e., $\mathbf{f}$), by solving the following sub-problem for the given $(\mathbf{t}, \eta, \mathbf{p}, \mathbf{b})$:
\vspace{-5mm}
%%%%%%%%%%%%%%%%%%
\begin{subequations}
\begin{align} \label{p3}
    \operatorname*{min}_{\mathbf{a},\mathbf{f}}\quad &\tau^{FL}\\
     \text{s.t.}\quad &a_kt_k^{up}b_k\log_2(1+\frac{p_k^{up}h_k}{b_k\mathcal{N}_0})\geq a_ks,\: \forall k,\label{a2_}\\
    &a_k\Big[t^{up}_kp_k^{up}+\zeta_kvC_kJ_k\log_2(\frac{1}{\eta})f^2_k\Big]\leq {{E}}_k^{\max}, \: \forall k,\label{p3-a2}\\
    & a_kt_H^{bc}{{B}}\log_2(1+\frac{p_H^{bc}h_k}{{{B}} \mathcal{N}_0})\geq a_ks, \: \forall k, \\
    & \sum_{k=1}^Ka_k\leq \frac{f_Ht^{cp}_H}{L_Hs},\\
    %&0\leq P\leq P^{\max}_{H},\\
    & \sum_{k=1}^{K}a_kb_k\leq {{B}}, \label{a4}\\
     &  a_k\in\{0,1\}, \: \forall k,\\
   % & t^{cp}_k+t^{up}_k\leq T,\\ 
    %& P_H^{cp}[n]t^{cp}_{H}[n]+P_H^{tr}[n]t^{tr}_{H}[n]\leq E_H[n], \: \forall n,\label{a6}\\
    %&\sum_{n=1}^N E_H[n]\leq E^{\max},\label{a7}\\
    % & K^{\min}_{th}\leq \sum_{k=1}^K a_k[n]\leq K^{\max}_{th}\leq K, \: \forall n,\label{a8}\\
   %  &  a_k[n]\in\{0,1\}, \: \forall n,k,\label{a8b}\\
      & f^{\min}_k\leq f_k\leq f^{\max}_k, \: \forall k.\label{p3-a5}
\end{align}
\label{af}
\end{subequations}
To minimize the delay at the given communication round, the local computing time needs to be minimized, which is equivalent to maximizing the local computation capability $\mathbf{f}$. Therefore, by using \eqref{p3-a2} and \eqref{p3-a5}, we can derive the optimal $\mathbf{f}^*=[f_k^*]_{1 \times K}$ as
\begin{align}
    f^*_k= \min\Bigg\{f_k^{\max}, \sqrt{\frac{a_k\left({{E}}_k^{\max}-t^{up}_kp^{up}_k\right)}{\zeta_kvC_kJ_k\log_2(\frac{1}{\eta})}}\Bigg\}, \: \forall k=1,\ldots,K.
\end{align}
%The optimal closed-form solution to this problem can be derived using the following Theorem~\ref{Thm1A}. 
%%%%%%%%%%%%%%%
%\begin{thm}\label{Thm1A}
%The optimal solution $(F^*,A^*)$ of problem %%\begin{align}
%       f^*_k[n]=\min\{f_k^{\max}, \sqrt{\frac{\mathbb{E}_k^{\max}-a_k[n]t^{up}_k[n]p_k[n]}{a_k[n]\zeta_kvC_kD_k\log_2(\frac{1}{\eta})}}\}, \: \forall k,n,
%\end{align}
%and
%\begin{align}
%    a_k^*[n]=\frac{-\mu_2+\sqrt{\mu_2^2-4\mu_1\mu_3}}{2\mu_1}, \: \forall k,n,
%\end{align}
%where $\mu_1$, $\mu_2$, and $\mu_3$ are the coefficients of the quadratic equation $\mu_1a_k^2[n]+\mu_2a_k[n]+\mu_3$, which is derived from KKT conditions.   
%\end{thm}%
%{\proof See Appendix~\ref{app:1A}.}

%We can notice that to minimize the local computation time and consequently the total delay, it is always efficient to utilize the maximum CPU computing capability as long as the local energy constraint is satisfied.
For a given $\mathbf{f}^*$, \eqref{af} is still an MINLP. To deal with it, we approximate the binary variable $a_k$ to a continuous variable such that $0\leq a_k\leq 1, \forall k$~\cite{cont-approx}. Subsequently, problem \eqref{af} becomes an LP with respect to $\mathbf{a}=[a_k]_{1 \times K}$. Hence, it can
be efficiently solved using standard convex optimization such as interior-point method.
%%%%%%%%%%%%%%%%%
\item{\textit{Optimal Power and Bandwidth Allocations $(\mathbf{p}^*, \mathbf{b}^*)$}}: To find the optimal client power and bandwidth allocation which minimizes the end-to-end uplink delay, we solve the following third sub-problem: 
\vspace{-3mm}
\begin{subequations}
\begin{align} \label{p4}
    \operatorname*{min}_{\mathbf{p}, \mathbf{b}} \quad &\tau^{UL}\\
     \text{s.t.}\quad & a_kb_k\log_2(1+\frac{p_k^{up}h_k}{b_k\mathcal{N}_0})\geq \frac{a_ks}{t_k^{up}},\: \forall k,\label{p4-const1}\\
     &p_k^{up}\leq \frac{a_k\Big[{{E}}_k^{\max}-\zeta_kvC_kJ_k\log_2(\frac{1}{\eta})f^2_k\Big]}{t^{up}_k}, \: \forall k,\label{p4-const2}\\
    %&0\leq P\leq P^{\max}_{H},\\
    & \sum_{k=1}^{K}a_kb_k\leq {{B}},  \label{p4-const3}
   % & t^{cp}_k+t^{up}_k\leq T,\\ 
    %& P_H^{cp}[n]t^{cp}_{H}[n]+P_H^{tr}[n]t^{tr}_{H}[n]\leq E_H[n], \: \forall n,\label{a6}\\
    %&\sum_{n=1}^N E_H[n]\leq E^{\max},\label{a7}\\
   %  & K^{\min}_{th}\leq \sum_{k=1}^K a_k[n]\leq K^{\max}_{th}\leq K, \: \forall n,\label{a8}\\
     %&  a_k[n]\in\{0,1\}, \: \forall n,k,\label{a8b}\\
      %& N\leq \frac{\alpha}{1-\eta}, \label{N}
      \\&0\leq p_k^{up}\leq p^{\max}_k, \: \forall k,\label{p4-const4}\\
    % & 0<\eta\leq 1, \label{a9}\\
     & b_k\geq 0, \: \forall k.
\end{align}
\label{Pr27}
\end{subequations}
The closed-form optimal solution of \eqref{Pr27} can be obtained using the following Theorem~\ref{Thm2}.
\begin{thm}\label{Thm2}
 The optimal solution $(\mathbf{p}^*=[p^{up^*}_k]_{1 \times K}, \mathbf{b}^*=[b_k^*]_{1\times K})$ of problem \eqref{Pr27} satisfies
\begin{align}
       & p_k^{up^*}=\min\left\{p_k^{\max}, \frac{a_k\Big[{{E}}_k^{\max}-\zeta_kvC_kJ_k\log_2(\frac{1}{\eta})f^2_k\Big]}{t^{up}_k}\right\}, \:\forall k,
\end{align}
and
\vspace{-5mm}
\begin{align}\label{user-b}
    &b_k^*=\frac{\beta_1+\beta_2(x_0-\pi_k)}{2\beta_2}+\frac{\sqrt{(\beta_1+\beta_2(x_0-\pi_k))^2+4\beta_2\pi_k(\beta_1+\beta_2x_0+1)}}{2\beta_2},\forall k,
\end{align}
where $\pi_k=\frac{a_kh_kp_k^{up^*}}{\mathcal{N}_0}$, $\beta_1=\ln(1+\frac{\pi_k}{x_0})$, $\beta_2=\frac{\pi_k}{x_0(x_0+\pi_k)}$, and $x_0$ is a real value related to Taylor expansion.  
\end{thm}
{\proof See Appendix~\ref{app:2}.}
%%%%%%%%%%%%%%%%%%%%%%

\item{\textit{Optimal HAPS resource allocation $(p_H^{bc^{*}},t_H^{bc^{*}},f_H^*)$}}: At the last step of the proposed iterative algorithm, we need to solve the following fourth sub-problem in order to optimize the HAPS resources allocation in the downlink:
\vspace{-3mm}
\begin{subequations}
\begin{align} \label{p5}
    \operatorname*{min}_{t_H^{bc}, p_H^{bc}, f_H}\quad & \tau^{DL}\\
     \text{s.t.}\quad
    %&0\leq P\leq P^{\max}_{H},\\
   % & t^{cp}_k+t^{up}_k\leq T,\\ 
    & t_H^{bc}\geq t_H^{\min}, \label{haps-1}\\
    &\zeta_Hf_H^2L_HQ_H+p_H^{bc}t_{H}^{bc}\leq {{E}}_H^{\max},\label{haps-2}
    %& N\geq \frac{\alpha}{1-\eta}, \\
     %  &\: 0\leq N\leq N^{\max},
\end{align}
\label{Pr30}
\end{subequations}
%\begin{align}
  %  b_k*[n]=\frac{p^*_k[n]h_k[n]}{\mathcal{N}_0(\sqrt{(\psi_na_k[n]\ln2)^2+2})}
%\end{align}
where
\vspace{-3mm}
\begin{align}
    t_H^{\min}=\frac{s}{\min_{\substack{k\in\mathcal{K} }} {{B}}\log_2(1+\frac{p_H^{bc}h_k}{{{B}}\mathcal{N}_0})}.
\end{align}
In this sub-problem, we assume that the parameters related to clients are given. We further decouple this problem into two new sub-problems to optimize $f_H$ and $(t_H^{bc},p_H^{bc})$ separately. Hence, for given HAPS broadcasting time and power allocation policies, \eqref{Pr30} is convex with respect to $f_H$. Following the Karush-Kuhn Tucker (KKT) conditions, the closed-form optimal computing capability of HAPS can be determined as
\vspace{-2.5mm}
\begin{align}\label{haps-f}
    f^*_H=\sqrt[3]{\frac{1}{2\omega\zeta_H},}
\end{align}
where $\omega$ is a Lagrangian multiplier associated with constraint~\eqref{haps-2} and is calculated in Appendix~\ref{app:3}.
%\\
%\indent 

Now, for given $f^*_H$, we solve the following problem to find the optimal HAPS transmit power and time allocations: 
\vspace{-3mm}
\begin{subequations}\label{main_haps}
\begin{align}\label{haps}
    \operatorname*{min}_{t_H^{bc}, p_H^{bc}}\quad & \tau^{DL}\\
     \text{s.t.}\quad
    %&0\leq P\leq P^{\max}_{H},\\
   % & t^{cp}_k+t^{up}_k\leq T,\\ 
    & t_H^{bc}\geq t_H^{\min}, \\
    & \zeta_H{f^*_H}^2L_HQ_H+p^{bc}_Ht^{bc}_{H}\leq {{E}}_H^{\max}.
\end{align}
\label{Pr33}
\end{subequations}
The closed-form solution for problem in \eqref{main_haps} can be obtained using the following Theorem~\ref{Thm3}:
\begin{thm}\label{Thm3}
 With given $f^*_H$, the optimal solution $(t_H^{bc^*},p_H^{bc^*})$ of problem~\eqref{Pr33} satisfies
 \vspace{-3mm}
 \begin{align}
    t^{bc^*}_H=t_H^{\min},
\end{align}
and
\vspace{-3mm}
\begin{align}\label{haps-p}
    p_H^{bc^*}=\sqrt{\frac{{{{B}}\mathcal{N}_0}}{\psi \min_{\substack{k\in\mathcal{K} }}h_{k}}},
\end{align}
where $\psi$ is the Lagrangian multiplier associated with constraint \eqref{haps-2}.
\end{thm}
\proof See Appendix~\ref{app:3}.\\
\end{enumerate}
\vspace{-2mm}
%\textcolor{blue}{\sout{It can be observed that all the sub-problems~\eqref{Pr21},~\eqref{af},~\eqref{Pr27}, and \eqref{Pr30} are convex, thus the KKT conditions are sufficient for optimality \cite[Section 5.5]{boyd}}.} 
The pseudo-code of the proposed iterative algorithm is summarized in Algorithm~\ref{alg2}.
%%%%%%
\vspace{-2mm}
\subsection{Convergence and Complexity Analysis}
\begin{enumerate}
    \item Convergence Analysis: The sub-problems~\eqref{Pr21},~\eqref{Pr27}, and \eqref{Pr30} are convex and sub-problem ~\eqref{af} becomes convex after the binary relaxation, w.r.t., to their respective block of optimization variables. Further, the objective functions in each sub-problem are lower-bounded by zero as client uplink time, communication delay, and HAPS downlink times cannot take a negative value. At each iteration $l$ in Algorithm~\ref{alg2}, due to convexity, the sub-problems give a unique minimum and always reduced objective function values from the previous iteration, i.e., $\tau^{FL, (l)}<\tau^{FL,(l-1)} < \ldots < \tau^{FL, (0)}$, where $\tau^{FL, (l)}$ represents the value of the objective function at iteration $l$ in Algorithm~\ref{alg2}.  The algorithm returns a monotonically decreasing sequence of objective function values lower bounded by zero, and thus, the sequence convergences. 
 %\textit{Proof:}
    \item Complexity Analysis: We provide the complexity analysis for Algorithm~\ref{alg2}. To solve the delay minimization problem~\eqref{Pr20} by using Algorithm~\ref{alg2}, the major complexity in each step lies in solving problem~\eqref{Pr21} and problem~\eqref{af}. To solve problem~\eqref{Pr21}, the complexity lies in obtaining the optimal $\eta^*$ and ${\bf{t^*}}$ according to Theorem~\ref{Thm2}, which involves complexity $\mathcal{O}(K \log_2 (1/\epsilon_1))$ with accuracy $\epsilon_1$. To solve problem~\eqref{af}, the overall complexity is dominated by the time complexity of solving the convex optimization problem to obtain ${\bf{a}}$ using the interior-point method. Therefore, the complexity of solving this problem is $\mathcal{O}(K^3)$. As a result, the total complexity of the proposed Algorithm~\ref{alg2} is $\mathcal{O}(l_{it}K^2(K+\log_2(1/\epsilon_1))$, where $l_{it}$ is the number of iterations for iteratively optimizing the variables in Algorithm~\ref{alg2}. Compared to the recent work~\cite{fl1}, our proposed approach demonstrates a lower computational complexity.
\end{enumerate}
\begin{figure*}[t!]
\begin{subfigure}[t]{0.5\textwidth}
\centering
    \includegraphics[width=80mm,scale=1]{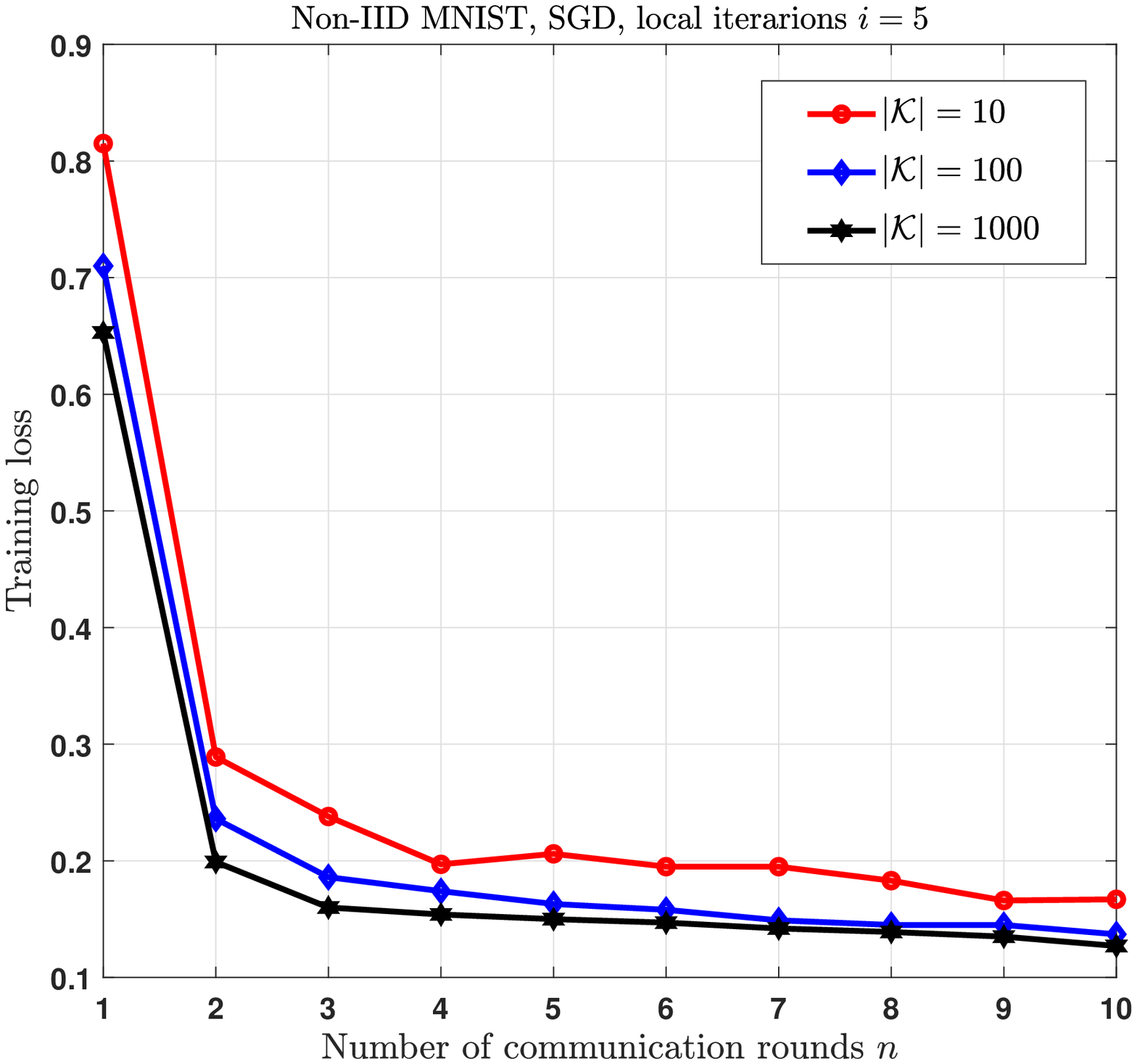}
   \caption{Training loss versus the number of communication rounds $n$. }
   \label{FL_acc_a}
   \end{subfigure}
  ~
   \begin{subfigure}[t]{0.5\textwidth}
   \centering
      \includegraphics[width=80mm,scale=1]{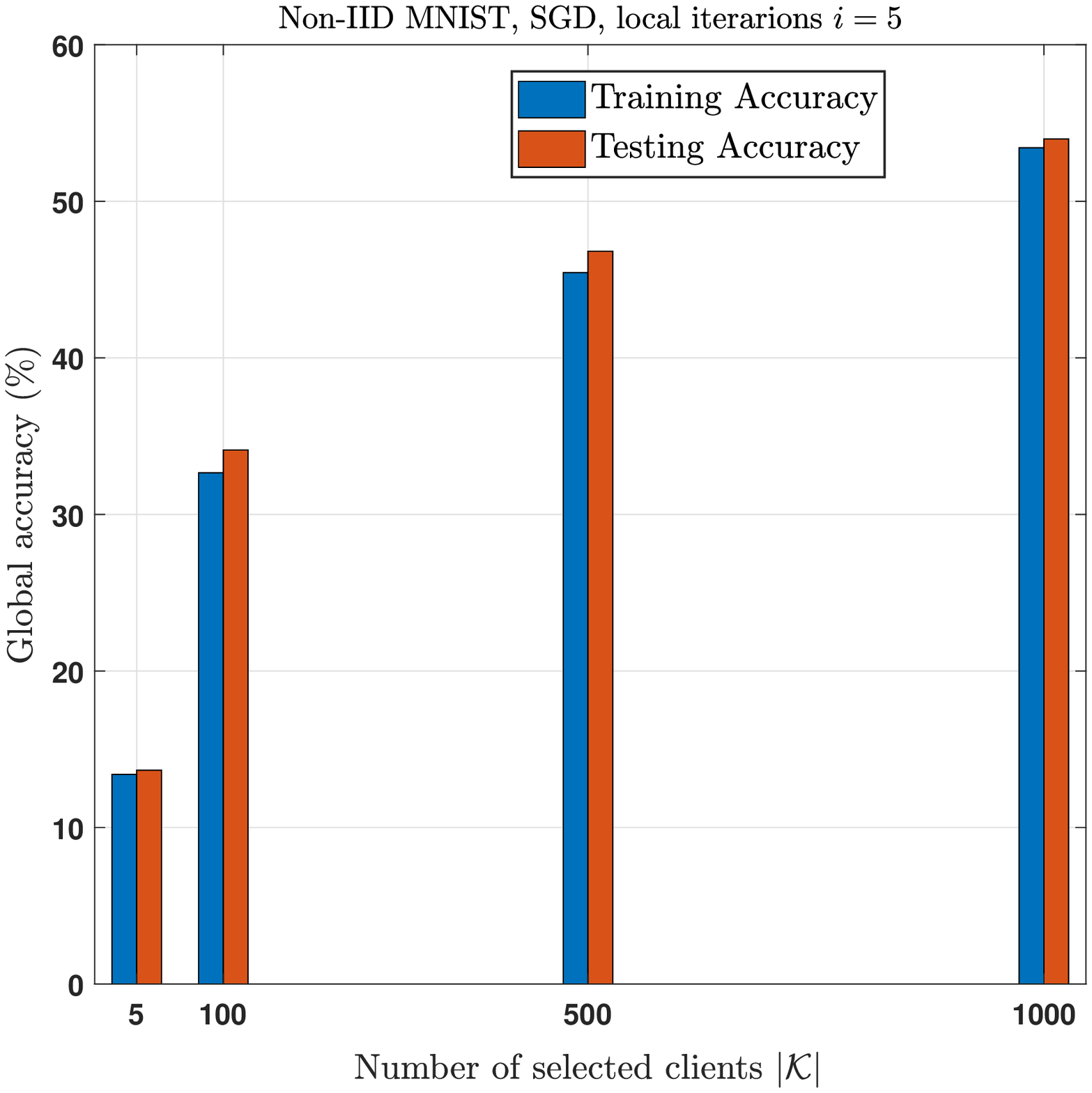}
  \caption{Global accuracy versus the number of selected clients $|\mathcal{K}|$.}
  \label{FL_acc_b}
    \end{subfigure}
    %\label{fig-acc}
    \caption{Global FL accuracy (i.e., $1-\epsilon$) and training loss for different number of selected clients $|\mathcal{K}|$.}
    \label{FL_acc}
    \vspace{-7mm}
\end{figure*}
%%%%
\vspace{-2mm}
\subsection{Summary of the Framework and Solution}
Our proposed FL framework, problem, and solution method are summarized as follows:
\begin{itemize}
    \item Our proposed FL framework, CCRA-FL, leverages a joint communication and computation resource optimization problem in a HAPS-enabled network architecture. According to Algorithm~\ref{alg1}, to solve the local optimization problem associated with the FL process, we adapt the FedAvg algorithm and use SGD to generate local updates on each client. CCRA-FL optimizes the FedAvg parameters by solving the network resource optimization problem subject to communication and computation constraints.
   % \item Using a HAPS-enabled network architecture, our proposed FL framework, CCRA-FL, optimizes communication and computation resources through Algorithm~\ref{alg1} and~\ref{alg2}. This process involves adapting the FedAvg algorithm, implementing SGD for local updates, and optimizing parameters within communication and computation constraints.
   \item In Theorem~\ref{conv-thm}, we present an upper bound on the convergence rate of the CCRA-FL algorithm, which demonstrates that a large number of accessible clients can be invited to the FL process through single-hop communications, leading to a higher FL convergence rate. However, we acknowledge the trade-off between communication efficiency and the FL convergence rate, which is why we focus on optimizing the client selection and  network resources while minimizing the end-to-end FL delay at every communication round.
    \item To solve the end-to-end delay minimization problem, we have developed an iterative algorithm, i.e., Algorithm~\ref{alg2}, that alternately optimizes four sub-problems until the algorithm converges to a given threshold. This problem is an MINLP, which cannot be optimally solved with standard convex optimization schemes. The proposed iterative algorithm optimizes HAPS and clients' resource allocation, local accuracy, and client selection policy.
\end{itemize}
\vspace{-3mm}
\subsection{Extension to Multi-HAPS Scenario}
The proposed FLSTRA framework for single HAPS can be extended to a multi-HAPS scenario for FL where  huge databases are located geographically far apart. Consider a scenario in which a network of HAPSs covers a set of unique geographical areas where each HAPS employs Algorithm 1 for FL. Each HAPS, after learning the area-specific global model, exchanges its model parameters with other HAPS to learn a final global model. To generate a final global model, HAPSs need to cooperate with each other such that all the area-specific global model parameters of HAPS must be aggregated. The aggregation can be performed in a decentralized fashion. As a result, a final global model is available in all geographical areas without sharing or compromising the huge databases. The FLSTRA can be extremely useful in applications such as self-driving vehicles, modern healthcare, and smart cities, where independent databases can be located in several large cities. To implement the FLSTRA for multi-HAPS scenarios, Algorithm 1 can be conveniently used at each HAPS. To arrive at a final global model, however, we need to develop an end-to-end decentralized, resource-aware inter-HAPS learning algorithm that minimizes delay.

%\textcolor{blue}{\sout{The proposed FL algorithm, initially designed for a single HAPS, adapts efficiently to this expanded setting. Specifically, the algorithm allocates resources among the HAPSs based on their geographical distribution and optimizes the selection of clients within each HAPS's coverage area. This adaptation ensures effective utilization of available resources, maximizes FL accuracy, and minimizes communication and computation costs. Furthermore, our joint communication and computation optimization problem, originally formulated for a single HAPS, can be extended for this multi-HAPS scenario. With multiple HAPSs involved, the iterative algorithm balances the resources across all HAPSs, effectively minimizing the end-to-end FL delay. Hence, our proposed algorithm and framework not only cater to a single HAPS scenario but also exhibit flexibility and robustness when extended to a more complex multiple HAPSs situation, thereby enhancing the efficacy of FL across a larger and more diversified geographical expanse.}}
%\vspace{3mm}
\section{Simulation Results And Discussion}\label{sim}
%\subsection{Total Delay minimization}
%\subsection{Convergence Behavior}
%\subsection{HAPS Displacement Impact}
In this section, we numerically evaluate and discuss the performance of the proposed CCRA-FL algorithm. For simulations, we consider $K\in\{10,100,1000\}$ clients uniformly distributed in a circular area with a radius of $50$ km, where a HAPS is located at its center at an altitude of $25$ km. We assume that the HAPS can drift from its original location with a variance from the range $[0.01,3]$ km. The average air-to-ground free-space path loss model in dB is given by \cite{path-loss}
\begin{equation}
 l_{k}=128.1 + 20 \log_{10} (d_k),   
\end{equation}
where $l_k$ and $d_k \in [25,55]$ km are the path loss and distance, respectively, for the communication link between the HAPS and client $k$. 
Moreover, the noise power spectral density is $\mathcal{N}_0 = -174$ dBm/Hz. We also assume that each client has $J_k = 500$ data samples, randomly selected
from the non-IID MNIST dataset~\cite{mnist}, with equal probability. The local computation capability parameter $C_k$ of client $k$ is uniformly distributed from the range $[1, 3] \times 10^4$ cycles/sample. Similarly, the HAPS's computation capability parameter $L_H$ is set to $3 \times 10^4$. The effective switched capacitance in local and HAPS computations are $\kappa = 10^{-28}$ and $\zeta_H=10^{-27}$, respectively. 
In addition, we set the uplink maximum transmission power $ p_k^{\max} = 10~\text{dBm}, \forall k$, while the maximum HAPS broadcasting power is set to $50$ dBm. The maximum computation capacity $f^{\max} =2$ GHz, the size of a local update is $s = 28.1$ kbits, and the total system bandwidth is set to $B = 20$ MHz.

To evaluate the proposed FL algorithm, we run our experiments on a non-IID MNIST dataset in a PyTorch implementation. In particular, the MNIST dataset is composed of $60,000$ training images, spread over $1200$ shards, each with $50$ images. The local batch size is set to $10$ images. Each client trains a convolutional neural network (CNN) that consists of two $5\times 5$ convolutional layers (the first with 32 channels, the second with 64, each followed by $2\times 2$ max pooling). The adopted loss function is the cross-entropy one. For comparison purposes, we use the following three baselines algorithms: 
 %%%%%%%%%%%%%%%%%%%%%%%
 %%%%
\begin{figure}[!t]
    \centerline{\includegraphics[width=95mm,scale=1]{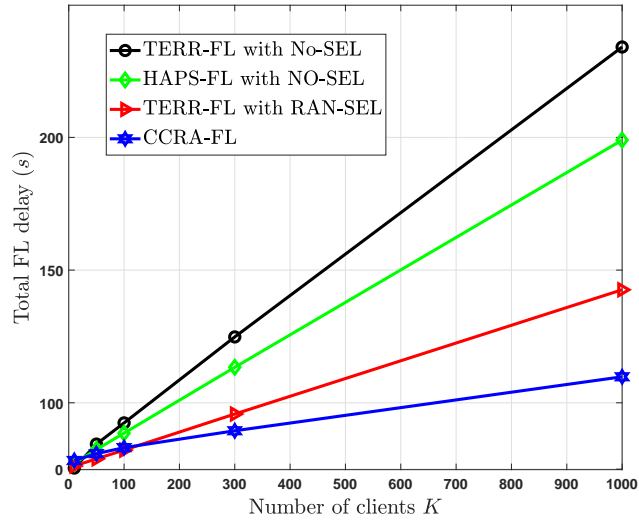}} 
    \caption{Total FL delay performance versus the number of clients, to achieve the same target training accuracy, compared with the benchmarks.}
    \vspace{-7mm}
    \label{bench}
\end{figure}
\begin{figure}[!t]
    \centerline{\includegraphics[width=92mm,scale=1]{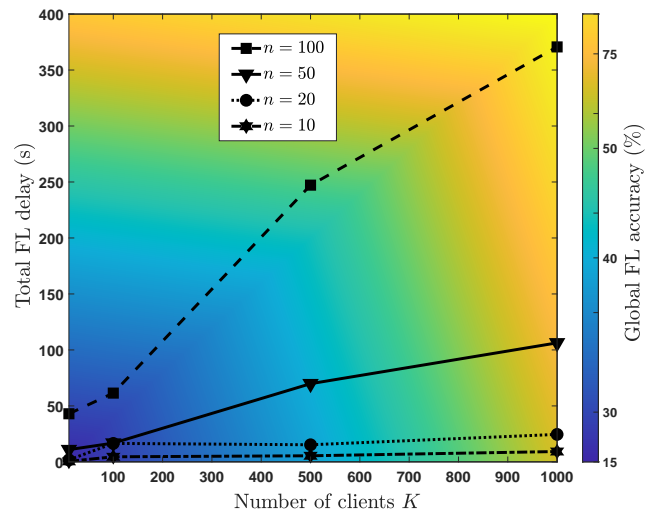}} 
    \caption{Total FL delay and global accuracy performance versus the number of clients for different communication rounds $n$.}
    \vspace{-7mm}
   \label{delay-k-N}
\end{figure}
%%%
%%%%
 \begin{figure}[!t]
    \centerline{\includegraphics[width=91mm,scale=1]{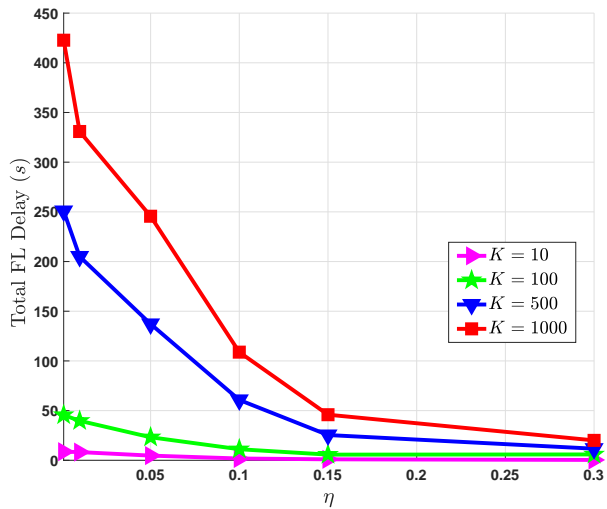}} 
    \caption{Total FL delay performance versus the local target accuracy $\eta$ for different number of clients $K$.}
    \vspace{-7mm}
    \label{fig2}
\end{figure}
\begin{figure}[!t]
    \centerline{\includegraphics[width=95mm,scale=1]{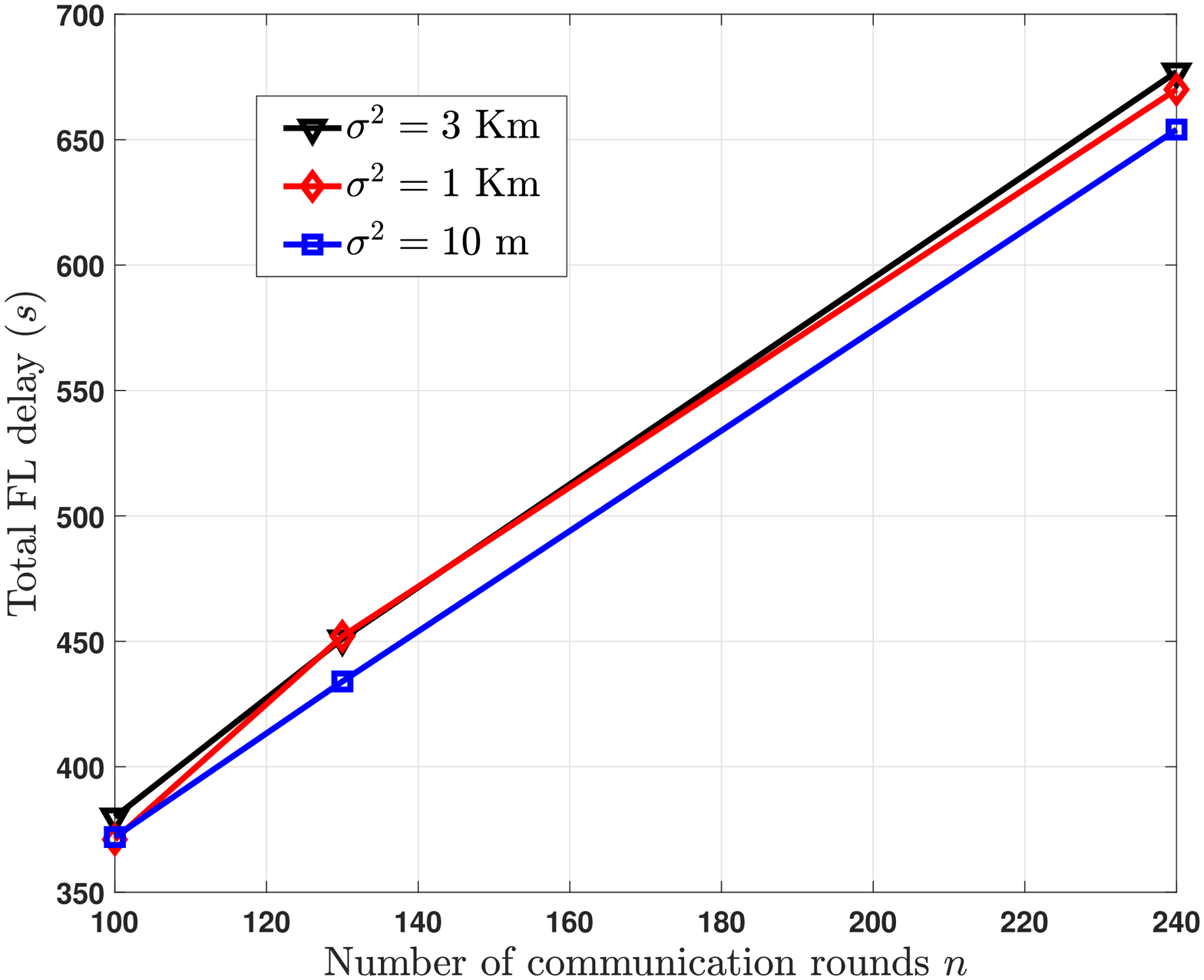}} 
    \caption{Total FL delay performance versus the number of communication rounds $n$ for different considerations of HAPS displacement variance $\sigma^2$ with $K=1000$.}
    \vspace{-7mm}
    \label{var-haps}
\end{figure}
%%%%
\begin{enumerate}
    \item TERR-FL with NO-SEL: In this algorithm, a client selection strategy is not employed (i.e., all clients are involved in learning) and the FL server is co-located with the terrestrial cloud server. 
    \item TERR-FL with RAN-SEL \cite{stndrd}: In this algorithm, the selection of clients is done randomly and each selected client receives an optimal allocation of radio resources. The term "radio resources" encompasses a range of available wireless communication resources, such as transmit power, bandwidth, and computational capacity, that are utilized by the clients to transmit their local model updates to the HAPS. The optimization of radio resources enables the clients to better utilize the available wireless resources and improves the efficiency of the FL process. Furthermore, it is important to note that in this algorithm, the FL server is co-located with the terrestrial cloud server, which further enhances the processing and computational capabilities of the system.
    \item HAPS-FL with NO-SEL: This algorithm considers no client selection strategy (i.e., all clients are involved in the training process). Also, the FL server is co-located with the HAPS.
\end{enumerate}

For the terrestrial FL setting, we consider a multi-hop scenario with $5$ MBSs each with a radius of up to $10$ km (urban area) to provide the same coverage that HAPS supports~\cite{stndrd}. These MBSs relay the local updates to a server at the cloud server for aggregation and model generation in the FL process. The path loss exponent is set to $4$. Further, under the terrestrial setting, the local updates reach the server at the MBS via two-hop communication.
%, which can be seen as a standard FL algorithm~\cite{stndrd}.  

In Fig.~\ref{FL_acc}, we evaluate the global accuracy and training loss performance of the proposed CCRA-FL algorithm by varying the numbers of selected clients, $|\mathcal{K}|$. The number of local iterations is fixed to $i=5$. In particular, Fig.~\ref{FL_acc_a} compares the FL training loss behaviour versus the number of communication rounds $n$ for three different considerations for the number of selected clients $|\mathcal{K}| \in  \{10, 100, 1000\}$. It can be observed that as the selected clients participation in the FL process increases, the accuracy of the CCRA-FL algorithm increases substantially up to $15\%$. Further, Fig.~\ref{FL_acc_b} investigates the FL accuracy behaviour with respect to the number of selected clients $|\mathcal{K}|$. It can be observed that the global training accuracy improves significantly up to $300\%$ with the number of clients participating. These results also corroborate with results in Theorem~\ref{conv-thm}.

In Fig.~\ref{bench}, we evaluate and compare the performance of the proposed CCRA-FL algorithm with the aforementioned benchmarks. It can be observed that to achieve the same global FL accuracy, the CCRA-FL algorithm results in less end-to-end FL delay, at a given communication round. This is due to the sub-optimal selection of FL parameters, i.e., $(n,|\mathcal{K}|, i)$, through solving the delay minimization problem. The increase in the delay trend of all algorithms is due to two main reasons. Firstly, as the client participation increases the HAPS server needs to process an increased number of data; thus, computing delay increases. Secondly, the allocated bandwidth among the clients decreases with the increasing number of selected clients. Consequently, the uploading rate reduces and the uploading time increases. However, the proposed algorithm for FLSTRA has the lowest increase in the delay trend. Specifically, compared with the TERR-FL with RAN-SEL algorithm, the proposed CCRA-FL algorithm shows a decrease up to $20\%$ in end-to-end FL delay behaviour due to the presence of LOS link and the single-hop communication between clients and the HAPS server.

Fig.~\ref{delay-k-N}, we investigate the dependency of end-to-end FL delay on the selection of pair $(n,|\mathcal{K}|)$ for a fixed number of local iterations $i=5$. It can be observed that the increase in the number of communication rounds $n$ can substantially influence the end-to-end FL delay over all the communication rounds and has a dominant impact among the FL parameters, i.e., $(n,|\mathcal{K}|, i)$. Moreover, it can be implied that to achieve a desired target accuracy, there exist a sub-optimal pair of the number of communication rounds $n$ and selected clients $|\mathcal{K}|$, and local iterations $i$ where the FL delay is minimized. For instance, to achieve $\epsilon=0.5$ with minimum FL delay at each communication round, the best $(n,|\mathcal{K}|)$ pair is $(10,500)$ for a fixed $i=5$.  
%%%%%%%%%%%%%%%%
\begin{table*}[!t]
 \caption{FLSTRA Total Delay, Training Loss, and Accuracy Performance.}
 \vspace{-5mm}
 \label{tab1}
\label{fl-performace}
\begin{center}
\begin{tabular}{p{2cm}|p{3cm}|p{2.4cm}|p{2.1cm}|p{2.1cm}|p{2.1cm}}
\toprule %
No. of selected clients ($|\mathcal{K}|$) & No. of communication rounds ($n$) & No. of local iterations ($i$) & Total FL delay ($s$) & Global training loss & Global accuracy $(1-\epsilon)\times 100$ \\ [.1\normalbaselineskip]\hline
$10$ & $10$ & $5$ &$0.458$& $7.50\%$&$15.33\%$  \\ [.1\normalbaselineskip]\hline
$10$& $10 %20
$&$50$ &$2.347$& $7.01 %6.60
\%$&$21.13 %22.50
\%$ \\ [.1\normalbaselineskip]\hline
$10$& $20$ & $5$&$3.72$& $6.14\%$&$32.27\%$ \\ [.1\normalbaselineskip]\hline
$100$& $10$& $5$ &$10.76$ & $2.19\%$&$29.84\%$ \\ [.1\normalbaselineskip]\hline
$100$& $10$& $50$&$17.61$&$1.80 \%$&$35.75\%$ \\ [.1\normalbaselineskip]\hline
$100$& $20$& $5$&$20.06$&$1.77 \%$&$50.03\%$ \\ [.1\normalbaselineskip]\hline
$1000$& $10$&$5$ &$23.51$& $1.61\%$&$45.37\%$ \\ [.1\normalbaselineskip]\hline
$1000$& $10$& $50$&$55.95$& $1.41\%$&$51.22\%$ \\ [.1\normalbaselineskip]\hline
$1000$& $20$& $5$&$61.43$& $1.38\%$&$60.73\%$ \\ [.1\normalbaselineskip]
\bottomrule 
\end{tabular}
\vspace{-7mm}
\end{center}
\end{table*}
%%%%%%%%%%%%%%%%
In Fig.~\ref{fig2}, we evaluate the end-to-end FL delay performance of the proposed CCRA-FL algorithm by varying the local accuracy $\eta$ for $K\in\{10, 100, 500, 1000\}$ and $n=20$. The results show that for the values of $\eta$ above $0.15$, the total end-to-end delay at a given communication round is less than $50$ seconds. Increasing the number of clients from $K=500$ to $K=1000$ with fixed $\eta=0.1$, i.e., higher local accuracy, it can be observed that the total FL delay increases by up to $38\%$ due to the increase in clients uploading time and HAPS computing time. % Further, for $K=1000$, which is possible due to HAPS large coverage, the FL accuracy is increased dramatically resulting increase in FL delay.
In Fig.~\ref{var-haps}, we evaluate the impact of undesirable displacement or drift of HAPS on the FL delay performance. It can be observed that as the displacement variance $\sigma^2$ increases, the total end-to-end delay increase up to $4\%$. In general, some selected clients experience bad channel quality in the current communication round due to displacement, thereby FL process takes a longer time to achieve the desired target accuracy over all communication rounds. However, the issue of bad channel quality is mitigated in the proposed CCRA-FL algorithm by employing the client selection strategy (i.e., step-2 in Algorithm~\ref{alg1}).
%however, using the proposed user selection strategy, the displacement on the FL delay performance, such that the increase is less than $4\%$ .}

Finally, Table~\ref{tab1} suggests an important and practical trade-off in the proposed FL setting: server can select more clients at each communication round while having each of them communicate less, and obtain the same accuracy as using fewer clients, but having each of them communicate more. The former may be preferable when many clients are available, but each has very limited upload bandwidth which is a setting common in practice.
%%%
%\begin{figure}[!t]
  %  \centerline{\includegraphics[width=87mm,s%cale=1]{HAPS_FL_10_mnist_cnn_10_C1_niid0.png}} 
  %  \caption{Total training Loss of benchmark (terrestrial setting) versus the number of global iterations $N$.}
  %  \label{fig-loss-N}
%\end{figure} 
%\begin{figure}[!t]
 %   \centerline{\includegraphics[width=87mm,scale=1]{HAPS_FL_500_mnist_cnn_10_C1_iid0.png}} 
  %  \caption{Total training loss of HAPS-enabled FL versus the number of global iterations $N$.}
  %  \label{fig-loss-N}
%\end{figure} 
%%%%%
%\begin{figure}[!t]
 %   \centerline{\includegraphics[width=30mm,scale=1]{.eps}} 
%    \caption{FL downlink delay performance versus HAPS maximum power $P^{\max}_H$ for different consideration of local accuracy $\eta$ ($K=10$).}
 %   \label{fig3}
%\end{figure}
%\begin{figure}[!t]
 %   \centerline{\includegraphics[width=30mm,scale=1]{.eps}} 
 %   \caption{FL downlink delay performance versus HAPS maximum power $P^{\max}_H$ for different consideration of local accuracy $\eta$ ($K=1000$).}
%    \label{fig4}
%\end{figure}
%%
%%
\vspace{-3mm}
\section{Conclusion}\label{concl}
 In this work, we introduced a novel idea of FLSTRA, which leverages the unique features of HAPS that allows large-scale FL. FLSTRA improves convergence rate and model accuracy but also increases delay. Hence, we developed a CCRA-FL algorithm with the aim to achieve the delay-accuracy trade-off. Particularly, we formulated a delay minimization problem to jointly optimize the communication and computation resource allocation (including uplink and downlink resources), client selection, and FL parameters. Since the formulated problem is non-convex, we decomposed it into four sub-problems and solved it in an iterative fashion. For some of the sub-problems, we also derived closed-form solutions. Our simulation results showed that to achieve the same level of model accuracy, the FLSTRA system significantly reduces the FL delay compared to the terrestrial benchmarks. Moreover, the proposed CCRA-FL algorithm for the FLSTRA is shown to be tolerant to HAPS undesirable displacement. Further, studying the impact of data distribution (i.e., non-IIDness level) on HAPS-enabled FL systems with more complex and larger datasets, such as CINIC-10 and CIFAR-10, and more sophisticated channel models is an interesting future direction. Additionally, considering multi-antenna devices as well as optimizing the 3D placement and beamforming of the HAPS server as part of the problem formulation will lead to improved performance of the FL system and reduction of the total end-to-end delay.
 %%%%%%%%%%%%%%%
 \vspace{-1mm}
\appendices
\section{Proof of Theorem~\ref{conv-thm}}
\setcounter{equation}{0}
\renewcommand{\theequation}{A.\arabic{equation}}
\label{app:1a}
Using the second-order Taylor expansion, we first rewrite $F(\mathbf{w}^{(n+1)})$ as
\vspace{-1mm}
\begin{align}
    F(\mathbf{w}^{(n+1)})&=F(\mathbf{w}^{(n)})+(\mathbf{w}^{(n+1)}-\mathbf{w}^{(n)})\nabla F(\mathbf{w}^{(n)}) +\frac{1}{2}(\mathbf{w}^{(n+1)}-\mathbf{w}^{(n)})^T\nabla^2F(\mathbf{w}^{(n)})(\mathbf{w}^{(n+1)}-\mathbf{w}^{(n)})\nonumber\\
    &\leq F(\mathbf{w}^{(n)})+(\mathbf{w}^{(n+1)}-\mathbf{w}^{(n)})^T\nabla F(\mathbf{w}^{(n)})+\frac{M}{2}\|\mathbf{w}^{(n+1)}-\mathbf{w}^{(n)}\|^2,
\end{align}
where the inequality stems from Assumption~\ref{assmp1}. According to the FL algorithm, at each communication round, the new global model is generated as $\mathbf{w}^{(n+1)}=\mathbf{w}^{(n)}+\frac{1}{|\mathcal{K}^{(n)}|}\sum_{k\in\mathcal{K}}\mathbf{z}_k^{(n)}$. Hence, we have
\vspace{-1mm}
\begin{align}
    F(\mathbf{w}^{(n+1)})\leq & \: F(\mathbf{w}^{(n)}) +\frac{1}{|\mathcal{K}^{(n)}|}\sum_{k\in\mathcal{K}}\nabla F(\mathbf{w}^{(n)})\mathbf{z}_k^{(n)} +\frac{M}{2|\mathcal{K}^{(n)}|^2}\Big\|\sum_{k\in\mathcal{K}}\mathbf{z}_k^{(n)}\Big\|^2,
\end{align}
According to the definition of the objective function of the local optimization problem in~\eqref{local}, we have
\vspace{-1mm}
\begin{align}
    F(\mathbf{w}^{(n+1)})&\nonumber\\ \leq & \: F(\mathbf{w}^{(n)}) +\frac{1}{\xi|\mathcal{K}^{(n)}|}\sum_{k\in\mathcal{K}^{(n)}}\Big[G_k(\mathbf{w}^{(n)},\mathbf{z}_k^{(n)}) F_k(\mathbf{w}^{(n)}+\mathbf{z}_k^{(n)})+\nabla F_k(\mathbf{w}^{(n)})\mathbf{z}_k^{(n)}\Big]+\frac{M}{2|\mathcal{K}^{(n)}|^2}\Big\|\sum_{k\in\mathcal{K}^{(n)}}\mathbf{z}_k^{(n)}\Big\|^2, 
\end{align}
Using Assumption~\ref{assmp1}, we have the following: 
\begin{align} \label{befr_tri}
    F(\mathbf{w}^{(n+1)}) \leq  F(\mathbf{w}^{(n)}) +\frac{1}{\xi|\mathcal{K}^{(n)}|}\sum_{k\in\mathcal{K}^{(n)}}\Big[G_k(\mathbf{w}^{(n)},\mathbf{z}_k^{(n)})-F_k(\mathbf{w}^{(n)})-\frac{u}{2}\|\mathbf{z}_k^{(n)}\|^2\Big]+\frac{M}{2|\mathcal{K}^{(n)}|^2}\Big\|\sum_{k\in\mathcal{K}^{(n)}}\mathbf{z}_k^{(n)}\Big\|^2. 
\end{align}
According to the triangle and mean inequalities, we have 
\begin{align} \label{tri}
    \Big\| \frac{1}{|\mathcal{K}^{(n)}|}\sum_{k\in\mathcal{K}^{(n)}} \mathbf{z}_k^{(n)}\Big\|^2 &\leq \Big[ \frac{1}{|\mathcal{K}^{(n)}|}\sum_{k\in\mathcal{K}^{(n)}}\|\mathbf{z}_k^{(n)}\| \Big]^2 \leq \frac{1}{|\mathcal{K}^{(n)}|}\sum_{k\in\mathcal{K}^{(n)}} \|\mathbf{z}_k^{(n)}\|^2. 
\end{align}
Substituting~\eqref{tri} into~\eqref{befr_tri}, we have
\begin{align}
    F(\mathbf{w}^{(n+1)})\leq & \: F(\mathbf{w}^{(n)}) +\frac{1}{\xi|\mathcal{K}^{(n)}|}\sum_{k\in\mathcal{K}^{(n)}}\Big[G_k(\mathbf{w}^{(n)},\mathbf{z}_k^{(n)})-F_k(\mathbf{w}^{(n)})-\frac{(u-M\xi)}{2}\|\mathbf{z}_k^{(n)}\|^2\Big],
\end{align}
According to~\eqref{local}, $F(\mathbf{w}^{(n)})=G_k(\mathbf{w}^{(n)},0)$. Hence, we have
\vspace{-5mm}
\begin{align} \label{A-derv}
    &F(\mathbf{w}^{(n+1)})\nonumber\\& \leq F(\mathbf{w}^{(n)}) +\frac{1}{\xi|\mathcal{K}^{(n)}|}\sum_{k\in\mathcal{K}^{(n)}}\Big[G_k(\mathbf{w}^{(n)},\mathbf{z}_k^{(n)})-G_k(\mathbf{w}^{(n)},\mathbf{z}_k^{(n)*})-(G_k(\mathbf{w}^{(n)},0)-G_k(\mathbf{w}^{(n)},\mathbf{z}_k^{(n)*}) )-\frac{(u-M\xi)}{2}\|\mathbf{z}_k^{(n)}\|^2\Big]\nonumber\\&\overset{\eqref{eta-bound}}\leq F(\mathbf{w}^{(n)}) -\frac{1}{\xi|\mathcal{K}^{(n)}|}\sum_{k\in\mathcal{K}^{(n)}}\Big[ (1-\eta)(G_k(\mathbf{w}^{(n)},0)- G_k(\mathbf{w}^{(n)},\mathbf{z}_k^{(n)*}))+\frac{(u-M\xi)}{2}\|\mathbf{z}_k^{(n)}\|^2\Big]\nonumber\\&\overset{\eqref{local}}=  F(\mathbf{w}^{(n)}) -\frac{1}{\xi|\mathcal{K}^{(n)}|}\sum_{k\in\mathcal{K}^{(n)}}\Big[ (1-\eta)\big(F_k(\mathbf{w}^{(n)}) \nonumber\\&\quad- F_k(\mathbf{w}^{(n)}+\mathbf{z}_k^{(n)*})+(\nabla F_k(\mathbf{w}^{(n)})-\xi\nabla F(\mathbf{w}^{(n)}))^T \mathbf{z}_k^{(n)*} \big) +\frac{(u-M\xi)}{2}\|\mathbf{z}_k^{(n)}\|^2\Big].
\end{align}
The optimal solution $\mathbf{z}_k^{(n)*}$ of problem~\eqref{local} always satisfies the first-order derivative condition, i.e., $\nabla G_k(\mathbf{w}^{(n)},\mathbf{z}_k^{(n)*})=0$. Therefore, we have
\vspace{-3mm}
\begin{align}\label{deriv}
    \nabla F_k(\mathbf{w}^{(n)}+\mathbf{z}_k^{(n)*})=\nabla F_k(\mathbf{w}^{(n)})-\xi\nabla F(\mathbf{w}^{(n)}).
\end{align}
Substituting~\eqref{deriv} into~\eqref{A-derv}, we have
\vspace{-3mm}
\begin{align}\label{A-derv2}
    F&(\mathbf{w}^{(n+1)}) \leq F(\mathbf{w}^{(n)}) +\frac{1}{\xi|\mathcal{K}^{(n)}|}\sum_{k\in\mathcal{K}^{(n)}}\Big[(1-\eta)\big(F_k(\mathbf{w}^{(n)}) \nonumber\\&\quad- F_k(\mathbf{w}^{(n)}+\mathbf{z}_k^{(n)*})+ \nabla F_k(\mathbf{w}^{(n)}+\mathbf{z}_k^{(n)*})^T \mathbf{z}_k^{(n)*}\big) +\frac{(u-M\xi)}{2}\|\mathbf{z}_k^{(n)}\|^2\Big].
\end{align}
Under Assumption~\ref{assmp1}, the following inequalities can be obtained~\cite{fl1}:
\vspace{-3mm}
\begin{align} \label{A-assmp1}
     F_k(\mathbf{w}^{(n)}) \geq &  F_k(\mathbf{w}^{(n)}+\mathbf{z}_k^{(n)*}) - \nabla F_k(\mathbf{w}^{(n)}+\mathbf{z}_k^{(n)*})^T \mathbf{z}_k^{(n)*} +\frac{u}{2}\|\mathbf{z}_k^{(n)*}\|^2,
\end{align}
and
\vspace{-5mm}
\begin{align} \label{A-assmp2}
    \|\mathbf{z}_k^{(n)*}\|^2 \leq \frac{1}{M^2} \|F_k(\mathbf{w}^{(n)}+\mathbf{z}_k^{(n)*}) -\nabla F_k(\mathbf{w}^{(n)})\|^2.
\end{align}
By applying~\eqref{A-assmp1} and \eqref{A-assmp2} to \eqref{A-derv2}, we can obtain
\vspace{-4mm}
\begin{align}
    &F(\mathbf{w}^{(n+1)})\nonumber\\& \leq F(\mathbf{w}^{(n)}) -\frac{1}{\xi|\mathcal{K}^{(n)}|}\sum_{k\in\mathcal{K}^{(n)}}\Big[\frac{(1-\eta)u}{2}\|\mathbf{z}_k^{(n)*}\|^2 +\frac{u-\xi M}{2}\|\mathbf{z}_k^{(n)}\|^2\Big]\nonumber\\&
    \leq F(\mathbf{w}^{(n)}) -\frac{1}{\xi|\mathcal{K}^{(n)}|}\sum_{k\in\mathcal{K}^{(n)}}\Big[\frac{(1-\eta)u}{2M^2}\| \nabla F_k(\mathbf{w}^{(n)}+\mathbf{z}_k^{(n)*})- \nabla F_k(\mathbf{w}^{(n)}) \|^2 +\frac{(u-M\xi)}{2}\|z_k^{(n)}\|^2\Big]\nonumber\\&
    \overset{\eqref{deriv}} = F(\mathbf{w}^{(n)}) -\frac{1}{\xi|\mathcal{K}^{(n)}|}\sum_{k\in\mathcal{K}^{(n)}}\Big[\frac{(1-\eta)u\xi^2}{2M^2}\| \nabla F(\mathbf{w}^{(n)}) \|^2  +\frac{(u-M\xi)}{2}\|\mathbf{z}_k^{(n)}\|^2\Big].
\end{align}
Based on Assumption~\ref{assmp1}, we can derive the following inequality~\cite{fl1}:
\vspace{-2mm}
\begin{align} \label{A-assmp3}
    \|\nabla F(\mathbf{w}^{(n)})\|^2 \leq u \Big[F(\mathbf{w}^{(n)})- F(\mathbf{w}^{*})\Big].
\end{align}
Applying~\eqref{A-assmp3}, we have
\vspace{-2mm}
\begin{align}
   F(\mathbf{w}^{(n+1)}) \leq & \: F(\mathbf{w}^{(n)})  - \frac{(1-\eta)u^2\xi}{2M^2}\Big[F(\mathbf{w}^{(n)}) - F(\mathbf{w}^*)\Big] - \frac{(u-M\xi)}{2\xi|\mathcal{K}^{(n)}|}\sum_{k\in\mathcal{K}^{(n)}}\|\mathbf{z}_k^{(n)}\|^2.
\end{align}
Accordingly, we can get
\vspace{-1mm}
{\normalsize\begin{align}
    & F(\mathbf{w}^{(n+1)}) - F(\mathbf{w}^*) \leq (1-\frac{(1-\eta)u^2\xi}{2M^2}) \Big[F(\mathbf{w}^{(n)}) - F(\mathbf{w}^*)\Big] - \frac{(u-M\xi)}{2\xi|\mathcal{K}^{(n)}|}\sum_{k\in\mathcal{K}^{(n)}}\|\mathbf{z}_k^{(n)}\|^2\nonumber\\&\leq (1-\frac{(1-\eta)u^2\xi}{2M^2})^{n+1} \Big[F(\mathbf{w}^{(0)}) - F(\mathbf{w}^*)\Big]- \frac{(u-M\xi)}{2\xi} \Big[ \frac{1}{|\mathcal{K}^{(0)}|}(1-\frac{(1-\eta)u^2\xi}{2M^2})^n \sum_{k\in\mathcal{K}^{(0)}}  \|\mathbf{z}_k^{(0)}\|^2 + \dots \nonumber\\&\quad + \frac{1}{|\mathcal{K}^{(n)}|}\sum_{k\in\mathcal{K}^{(n)}} \|\mathbf{z}_k^{(n)}\|^2 \Big]\nonumber\\&=
    (1-\frac{(1-\eta)u^2\xi}{2M^2})^{n+1} \Big[ F(\mathbf{w}^{(0)}) - F(\mathbf{w}^*)\Big]  +\frac{(M\xi-u)}{2\xi}\Big[ \sum_{n^{\prime}=0}^n  \frac{1}{|\mathcal{K}^{(n^\prime)}|}(1-\frac{(1-\eta)u^2\xi}{2M^2})^{n-n^\prime}\sum_{k\in\mathcal{K}^{(n^{\prime})}}\|\mathbf{z}_k^{(n^\prime)}\|^2\Big].
\end{align}}

To ensure that the convergence upper bound is always positive, we need to ensure that both terms on the left-hand side of the bound are positive. It is known that the loss function for an optimal global model is always less than that for any other models, which means that $F(\mathbf{w}^{(0)}) - F(\mathbf{w}^*)\geq 0$. Furthermore, from equation~\eqref{fl-bound1}, we can see that the local target accuracy is bounded as $0\leq \eta \leq 1$, which implies that $0\leq 1-\eta\leq 1$. By combining these two facts, we obtain the inequality $0\leq \frac{(1-\eta)u^2\zeta}{2M^2}\leq \frac{u^2\zeta}{2M^2}$.
For the second term, we require that $M\zeta -u\geq 0$. Therefore, by considering these two conditions, we can obtain the following condition on $u$ and $M$, which are both positive values:
\vspace{-2mm}
\begin{align}
    u\leq M\min\Big\{\zeta,\sqrt{\frac{2}{\zeta}}\Big\}.
\end{align}
%It is important to ensure that this condition is satisfied in order to guarantee a positive convergence upper bound, which is essential for the successful operation of the proposed FL system. By carefully selecting appropriate values for $u$ and $M$ based on this condition, we can ensure that the FL system will converge in a reasonable amount of time while maintaining a high level of accuracy.

This completes the proof.
%%%%%%%%%
\vspace{-4mm}
\section{Proof of Theorem~\ref{Thm1}}
\setcounter{equation}{0}
\renewcommand{\theequation}{A.\arabic{equation}}
\label{app:1}
To minimize the delay $\tau^{FL}$, the local uploading time $t^{up}_k$ needs to be minimized. Hence, to minimize the $t^{up}_k$ from~\eqref{p2-const2}, we have
\vspace{-6mm}
\begin{align}
    t^{up^*}_k= t^{\min}_k, \: \forall k.
\end{align}
Given this solution, \eqref{p2} is a convex problem with respect to local accuracy $\eta$. Let $\mathcal{L}_1(\eta, \boldsymbol{\lambda}, \theta)$ denote the Lagrangian function which can be defined as
\vspace{-2mm}
\begin{align}
    &\mathcal{L}_1(\eta, \boldsymbol{\lambda}, \theta)=  \operatorname*{max}_k  a_k(t_k^{up}+\frac{vC_kJ_k\log_2(\frac{1}{\eta})}{f_k}) +\sum_{k=1}^K \lambda_{k}\Big[ a_k\big(t^{up}_kp_k^{up}+\zeta_kvC_kJ_k\log_2(\frac{1}{\eta})f^2_k\big)
    -{{E}}^{\max}_k \Big] +\theta(\eta-1),
\end{align}
where $\boldsymbol{\lambda}=[\lambda_k]_{1 \times K}$ and $\theta$ are Lagrangian multipliers associated with constraints~\eqref{p2-const2} and \eqref{p2-const3}, respectively.  
Using the KKT conditions, i.e.,  $\frac{\partial \mathcal{L}_1}{\partial \eta}=0$, the closed-formed solution~\eqref{eta} can be derived.
%\section{Proof of Theorem~\ref{Thm1A}}
%\setcounter{equation}{0}
%\renewcommand{\theequation}{B.\arabic{equation}}
%\label{app:1A}
%To minimize the total delay, the local computing time needs to be minimized which is equivalent to maximizing the computation capability $F$. Therefore, by using \eqref{p3-a2} and \eqref{p3-a5}, we can derive the optimal $F^*$ as
%\begin{align}
  %  f^*_k[n]= \min\{f_k^{\max}, \sqrt{\frac{\mathbb{E}_k^{\max}-a_k[n]t^{up}_k[n]p_k[n]}{a_k[n]\zeta_kvC_kD_k\log_2(\frac{1}{\eta})}}\}, \: \forall k,n.
%\end{align}
%Given $F^*$, \eqref{p3} is still a mixed-inter programming problem which cannot be solved using convex optimization framework. To deal with the non-convexity, we can approximate the integer variable $a_k[n]$ with a continuous variable such that $0\leq a_k[n]\leq 1$~\cite{}. It is observed that problem \eqref{p2} is convex with respect to $A$ and can be solved using Lagrangian method. Let $\mathcal{L}_2$ denote the corresponding Lagrangian function defined as
%\begin{align}
 %   \mathcal{L}_2()
%\end{align}
%%%%%%%%%%%%%%
\vspace{-4mm}
\section{Proof of Theorem~\ref{Thm2}}
\setcounter{equation}{0}
\renewcommand{\theequation}{C.\arabic{equation}}
\label{app:2}
\indent From a delay minimization perspective, local uploading power $p_k^{up}$ at each communication round  needs to be maximized while the local energy constraint is satisfied. From $\eqref{p4-const2}$ and $\eqref{p4-const4}$, $p_k^{up}$ is optimized as
\vspace{-2mm}
\begin{align}
 & p_k^{up^*}=\min\{p_k^{\max}, \frac{a_k\big[{{E}}_k^{\max}-\zeta_kvC_kJ_k\log_2(\frac{1}{\eta})f^2_k\big]}{t^{up}_k}\}, \:\forall, k.
\end{align}

With given $\mathbf{p}=[p_k^{up^*}]_{1\times K}$, problem~\eqref{p4} is convex with respect to $b_k$ and can be solved using the Lagrangian method. Let $\mathcal{L}_2(\mathbf{b},\boldsymbol{\gamma},\psi)$ denote the Lagrangian function which can be defined as
\vspace{-2mm}
\begin{align}
    \mathcal{L}_2(\mathbf{b},\boldsymbol{\gamma},\psi)= \tau^{FL}-\sum_{k=1}^K\gamma_{k}&\Big[a_kb_k\log_2(1+\frac{h_kp^{up^*}_k}{b_kN_0})-\frac{a_ks}{t^{up}_k}\Big]+\psi(\sum_{k=1}^K a_k b_k-{{B}}),
\end{align}
where ${\boldsymbol{\gamma}}=[\gamma_k]_{1\times K}$ and $\psi$ are Lagrangian multipliers associated with constraints \eqref{p4-const1} and \eqref{p4-const3}, respectively.
Following KKT conditions, we then need to solve the following problem:
\vspace{-1mm}
\begin{align}\label{app2-kkt}
\frac{\partial \mathcal{L}_2}{\partial b_k}=\gamma_{k}\Big[a_k\log_2(1+\frac{\pi_k}{b_k})&-\frac{a_k\pi_k}{b_k(1+\frac{\pi_k}{b_k})\ln 2}\Big]-\psi_na_k=0, \: \forall k,
\end{align}
where $\pi_k=\frac{h_kp_k^{up^*}}{\mathcal{N}_0}$. \eqref{app2-kkt} is hard to be solved analytically in this form. Hence, by using first-order Taylor expansion at point $x_0\geq \pi_k, \: \forall k\in\mathcal{K}$, we have
\vspace{-3mm}
\begin{align}
\beta_1-\beta_2(b_k-x_0)+\frac{\pi_k}{b_k+\pi_k}-\frac{\psi\ln 2}{\gamma_{k}}=0,
\end{align}
where $\beta_1=\ln(1+\frac{\pi_k}{x_0})$ and $\beta_2=\frac{\pi_k}{x_0(x_0+\pi_k)}$. With $\psi=0$, the solution to this quadratic equation can be obtained as \eqref{user-b}.
\vspace{-4mm}
%\begin{align}
%    &\frac{\partial \mathcal{L}}{\partial B}=0,\\
 %   & \sum_{n=1}^N\sum_{k=1}^K\gamma_{k,n}(b_k[n]\log_2(1+\frac{h_k[n]p^*_k[n]}{b_k[n]N_0})-s)=0,\\
 %   & \sum_{n=1}^N\mu_{n}(\mathbb{B}-\sum_{k=1}^Ka_k[n]b_k[n])=0.
%\end{align}
%Then, by using first-order Taylor expansion at point $x_0$, we have
\vspace{-4mm}
\section{Proof of Theorem~\ref{Thm3}}
\setcounter{equation}{0}
\renewcommand{\theequation}{D.\arabic{equation}}
\label{app:3}
From \eqref{p5}, we can notice that problem is convex with respect to $f_H$ since the second-order derivatives are positive. Hence, we can apply the Lagrangian method to find the optimal solution. Let $\mathcal{L}_3(f_H,\omega)$ denote the Lagrangian function which is defined as
\vspace{-3mm}
\begin{align}
    \mathcal{L}_3&(f_H,\omega)= (\frac{L_H\sum_{k=1}^Ka_k}{f_H} + t_H^{bc})+ \omega \Big[ (\zeta_Hf_H^2L_H\sum_{k=1}^Ka_k+p_H^{bc}t_{H}^{bc})-{{E}}_H^{\max}\Big].
\end{align}
Then, by solving $\frac{\partial \mathcal{L}_3}{\partial f_H}=0$, the optimal HAPS computing capability $f_H^*$ can be derived as~\eqref{haps-f}.%Moreover, the Lagrangian multiplier can be updated as...\\
\indent With given $f_H^*$, \eqref{p5} can be solved to find optimal HAPS broadcasting time and power at each communication round. From a delay minimize perspective, it is always efficient to allocate the minimum time for broadcasting the FL global update at HAPS. Hence, to minimize the broadcasting time from \eqref{p5}, we have
\vspace{-3mm}
\begin{align}
    t_H^{bc^*}=\frac{s}{\min_{\substack{k\in\mathcal{K}}} {{B}}\log_2(1+\frac{p_H^{bc}h_k}{{{B}}\mathcal{N}_0})}.
\end{align}

It can be verified that \eqref{p5} is a convex problem with respect to $p_H^{bc}$ and then can be solved using the Lagrangian multipliers method. Let $\mathcal{L}_4(p_H^{bc},\psi)$ denote the Lagrangian function defined as
\begin{align}
    &\mathcal{L}_4(p_H^{bc},\psi) =\frac{L_H\sum_{k=1}^Ka_k}{f_H} + \frac{s}{{{B}}\log_2(1+\frac{p_H^{bc}\min_{\substack{k\in\mathcal{K}}}h_k}{{{B}}\mathcal{N}_0})} +\psi\Big[(\zeta_Hf_H^2L_H\sum_{k=1}^Ka_k[n]+p_H^{bc}t_{H}^{bc})-{{E}}_H^{\max} \Big],
\end{align}
where $\psi$ is a Lagrangian multiplier.
%Let $k^\prime$ be the solution to $\min_k a_k[n]h_k[n]$ such that $k^\prime\in\{1,\dots,K\}$.
Note that since $\log$ function is increasing with respect to $p_H^{bc}$, then we have $\min_{\substack{k\in\mathcal{K}}}{{B}}\log_2(1+\frac{p_H^{bc}h_k}{{{B}}\mathcal{N}_0})= {{B}}\log_2(1+\frac{p_H^{bc}\min_{\substack{k\in\mathcal{K}}} h_k}{{{B}}\mathcal{N}_0})$. Then by applying KKT conditions, we can obtain the closed-form solution~\eqref{haps-p}.  
 %%%%%%%%%%%%%%%%%%%%%%%%%%%%%%%%%%%%%
 
\end{document}